\documentclass[acmsmall]{acmart}
\usepackage{xspace}
\usepackage[normalem]{ulem} 
\AtBeginDocument{%
  }

\setcopyright{cc}
\setcctype{by-nc-nd}
\acmJournal{TORS}
\acmYear{2026} \acmVolume{4} \acmNumber{1} \acmArticle{46}
\acmMonth{3} \acmPrice{} \acmDOI{10.1145/3795792}

\usepackage{todonotes}
\usepackage[inline]{enumitem}
\presetkeys%
    {todonotes}%
    {inline,backgroundcolor=yellow}{}

\usepackage{xspace}
\newcommand{\palpha}{\ensuremath{\text{P}^3\!\alpha}\xspace}
\newcommand{\pbeta}{\ensuremath{\text{RP}^3\!\beta}\xspace}
\newcommand{\iALS}{iALS\xspace}
\newcommand{\MFBPR}{MF-BPR\xspace}
\newcommand{\EASER}{EASE$^R$\xspace}
\newcommand{\idest}{i.e.,\xspace}
\newcommand{\eg}{e.g.,\xspace}
\usepackage{dirtytalk}
\usepackage[normalem]{ulem}
\usepackage{multirow}

\usepackage{xcolor}
\definecolor{badcol}{RGB}{255,224,224}   
\definecolor{badcol2}{RGB}{255,200,200}  
\usepackage{tikz}
\usetikzlibrary{positioning,fit,backgrounds,decorations.pathreplacing,calc,arrows.meta}

\newcommand{\tableArticleResult}[7]{
\begin{table}[th!]
    \caption{Results on the #4 dataset. Bold values indicate the top scores among the baselines and the mean of our #2 results, while underlined values denote the best scores among the baselines and the results of #2 reported in the original paper. #6}
    \ifnum#7=0
        \label{tab:#2-#3-result}
    \else
        \label{tab:#2-#3-result-appendix}
    \fi

    \footnotesize
    \centering

    \ifnum#5=0
        \ifnum#7=0
            \resizebox{\linewidth}{!}{\input{tables/#1_#3_article_metrics_latex_results.txt}}
        \else
            \resizebox{\linewidth}{!}{
            \input{tables/appendix_outcomes/#1_#3_article_metrics_latex_results.txt}
            }
        \fi
    \else
        \ifnum#7=0
            \input{tables/#1_#3_article_metrics_latex_results.txt}
        \else
            \input{tables/appendix_outcomes/#1_#3_article_metrics_latex_results.txt}
        \fi
    \fi
\end{table}
}

\newcommand{\hyperparamsArticleTable}[4]{
\begin{table}[th!]
    \caption{#2's hyperparameter values used in the reproducibility experiments #4}
    \label{tab:#2-hyperparams}
    \footnotesize
    \centering

    \ifnum#3=0
        \resizebox{\linewidth}{!}{\input{tables/appendix/#1_hyperparams.txt}}
    \else
        \input{tables/appendix/#1_hyperparams.txt}
    \fi
\end{table}
}

\begin{document}

\title[Diffusion Recommender Models and the Illusion of Progress]{Diffusion Recommender Models and the Illusion of Progress: A Concerning Study of Reproducibility and a Conceptual Mismatch}

\author{Michael Benigni}
\orcid{0009-0002-2810-5800}
\affiliation{%
    \department{ReMAP Lab}
    \institution{Politecnico di Milano}
    \country{Italy}
    \city{Milano}
}
\email{michael.benigni@polimi.it}

\author{Maurizio {Ferrari Dacrema}}
\orcid{0000-0001-7103-2788}
\affiliation{%
    \department{ReMAP Lab}
    \institution{Politecnico di Milano}
    \country{Italy}
    \city{Milano}
}
\email{maurizio.ferrari@polimi.it}

\author{Dietmar Jannach}
\orcid{0000-0002-4698-8507}
\affiliation{%
  \institution{University of Klagenfurt}
  \country{Austria}
    \city{Klagenfurt}
}
\email{dietmar.jannach@aau.at}

\begin{abstract}
Countless new machine learning models are published every year and are reported to significantly advance the state-of-the-art in \emph{top-n} recommendation. However, earlier reproducibility studies indicate that progress in this area may be quite limited, due to widespread methodological issues, \eg comparisons with untuned baseline models, creating an \emph{illusion of progress}. In this work, we examine whether these problems persist in today's research by attempting to reproduce nine SIGIR 2023 and 2024 recommendation algorithms based on Denoising Diffusion Probabilistic Models, a recent but rapidly expanding research area. Only 25\% of reported results are fully reproducible and, since the original papers relied on \emph{weak baselines}, they do not establish the superiority of diffusion models over state-of-the-art methods. In our controlled evaluations, well-tuned simpler baselines consistently exceed the diffusion-based models' effectiveness reported in the original papers.
Furthermore, we identify key mismatches between the characteristics of diffusion models and those of the traditional \emph{top-n} recommendation task, raising doubts about their suitability for recommendation. Moreover, in the analyzed papers, the generative capabilities of these models are constrained to a minimum.
Overall, our results call for greater scientific rigor and a disruptive change in the research and publication culture in this area.
\end{abstract}

\ccsdesc[500]{Information systems~Recommender systems}
\ccsdesc[300]{Information systems~Collaborative filtering}
\ccsdesc[200]{General and reference~Evaluation\vspace{-1mm}}

\keywords{Recommender Systems, Evaluation, Reproducibility, Diffusion Models}
\maketitle

\section{Introduction}
\label{sec:introduction}
Research in the area of recommender systems is flourishing, with a constant stream of new machine learning models proposed each year to more accurately predict user preferences from logged interaction data. One main driver of algorithmic research in recommender systems is the development of new machine learning models or deep learning architectures in other areas of applied machine learning. Examples include the application of recurrent neural networks, convolutional neural networks, graph neural networks, attention mechanisms, and transformer-based architectures to recommendation problems~\cite{DBLP:journals/corr/HidasiKBT15,kang2018self,sun2019bert4rec,Wang2019GraphCNN,Wu2022GNN}. Most recently, various researchers have investigated the value of incorporating principles of \emph{Denoising Diffusion Probabilistic Models} (DDPMs)~\cite{DBLP:conf/icml/Sohl-DicksteinW15,Ho2020Diffusion} into neural recommendation architectures~\cite{wei2025diffusionmodelsrecommendationsystems,Walker2022Codigem,DBLP:conf/sigir/WangXFL0C23,DBLP:conf/sigir/HouPS24,DBLP:conf/sigir/ZhuWZX24,DBLP:conf/sigir/ZhaoWXSFC24}. This constitutes a recent but rapidly growing research direction, with applications in collaborative filtering, sequential recommendation, and various multimodal tasks, from image generation to text-to-recommendation, among others \cite{wei2025diffusionmodelsrecommendationsystems}. Given the success of DDPMs in other fields and their rapid adoption in recommendation, we consider it timely to conduct a reproducibility study. Furthermore, none of the papers we analyze has conducted an in-depth discussion to motivate why DDPMs, which were designed for image generation tasks, should be the right architecture for tasks such as collaborative filtering, which rely on very different assumptions. As we will discuss, there are strong mismatches between DDPMs and traditional collaborative filtering tasks that have been underestimated and, in our view, constitute fundamental obstacles to the application of DDPMs.

Denoising Diffusion Probabilistic Models are a generative neural architecture that represents the state-of-the-art in image synthesis, \idest the generation of artificial images or the enhancement of existing ones from inputs like textual descriptions. Technically, these models learn to reverse a stepwise noise-adding process that is progressively applied to the data. In doing so, they are able to model complex distributions by decomposing the problem into simpler subtasks. Wang et al.~\cite{DBLP:conf/sigir/WangXFL0C23} established\footnote{An earlier model that incorporates ideas from DDPMs into collaborative filtering was presented in~\cite{Walker2022Codigem}.} the relationship between DDPMs and recommender systems, stating that recommender systems \emph{``essentially infer the future interaction probabilities based on corrupted historical interactions.''}
By ``corruption,'' the authors refer to the noise that can commonly be found in recorded user-item interactions~\cite{Wang2021Noise}. They then proposed the \emph{DiffRec} model, which adapts the principles of DDPMs to the recommendation task, and successfully benchmarked it against various state-of-the-art models. Subsequent works then proposed alternative ways of incorporating DDPMs into recommendation models, reporting effectiveness even superior to \emph{DiffRec}~\cite{DBLP:conf/sigir/HouPS24,DBLP:conf/sigir/ZhuWZX24,DBLP:conf/sigir/ZhaoWXSFC24}.\footnote{For a survey of the various uses of DDPMs in recommender systems, see~\cite{wei2025diffusionmodelsrecommendationsystems}.}

Overall, these findings suggest that DDPMs represent a promising direction for further advancing the field of recommender systems, and that, as a community, we are continuing to make progress in increasing the prediction and ranking accuracy of our recommendation models. However, independently of specific architectures like DDPMs, 
one may ask how it is possible that probably hundreds of algorithmic papers are published each year, with almost all claiming to improve prediction accuracy over previous models by a few percent. In fact, similar questions were raised more than fifteen years ago in the broader field of information retrieval. In that area, a study by Armstrong et al.~\cite{Armstrong:2009:IDA:1645953.1646031} published in 2009 revealed that the reported improvements for a common ad-hoc retrieval task ``don't add up'' over the course of the previous ten years. One main reason for this illusion of continued progress was found in the consideration of weak baselines in experimental evaluations.

Similar problems have been identified in the area of recommender systems during the past decade as well. Beyond the choice of weak baselines, it was also found that a lack of proper tuning of existing, and potentially actually strong, baselines may lead to a phenomenon that was referred to as `phantom progress' in~\cite{ferraridacremaetal2019}. Reproducibility studies in~\cite{ferraridacremaetal2019,ferraridacrema2020tois,Rendle2020NeuMF,DBLP:conf/sigir/ShehzadDJ25} showed that sometimes even decade-old and conceptually quite simple methods, when properly tuned, are almost consistently more effective than the recent deep learning models of the time for \emph{top-n} recommendation tasks. Later research then re-assessed the effectiveness of widely-used and more than fifteen years old matrix factorization models, finding that they are still competitive with models that would be considered the state-of-the-art today~\cite{RendleKZK22Revisiting,RevisitingBPR2024}. Similar findings were also reported for \emph{sequential} recommendation models~\cite{ludewiglatifiumuai2020}, models based on architectures like Graph Neural Networks (GNNs)~\cite{shehzad2024performance,Anelli2023Myth,TOIS_SIGIR_22}, and even in other areas of applied machine learning like time-series forecasting~\cite{Makridakis2018}.
In Section~\ref{sec:previous-studies}, we provide an overview of known methodological issues in recommender systems research, and we review multiple previous studies which had identified issues that are similar to those found in our present work.

Since such 
methodological issues have been well known for several years, we were wondering if these problems still exist today. With this work, we therefore aimed to \emph{(i)} attempt to reproduce the latest findings in recommender systems research based on DDPMs, to \emph{(ii)} check for the existence of potential methodological issues, and to \emph{(iii)} assess the effectiveness of DDPM-based models relative to a number of properly tuned baselines of different families. Furthermore, we reflect on the theoretical foundations of DDPMs and highlight key mismatches with the personalized recommendation task, raising concerns regarding their suitability for this domain.

From a methodological perspective, we closely follow earlier studies like~\cite{ferraridacremaetal2019,ferraridacrema2020tois}, where the artifacts provided by the authors of the original papers are used as a basis for reproducibility and benchmarking. This way, we ensure that the original evaluation procedures are followed as closely as possible. As a basis for our study, we considered four papers proposing nine different recommendation models based on DDPMs that were published in 2023 and 2024 in the proceedings of the top-ranked ACM SIGIR conference series.

Overall, our findings are disillusioning. Our analyses revealed that a number of worrying methodological practices are still common in the examined papers and that reproducing existing works remains challenging.  One notable aspect in this context is that some of the examined DDPM-based recommendation models can exhibit substantial variance across several execution runs, with their effectiveness varying between runs by as much as  18\%.
Furthermore, our experimental evaluations showed that, for all examined papers, there was at least one baseline model that was equally or more effective than the proposed DDPM-based model. In several cases, even simple neighborhood-based methods were more effective than the latest approaches based on DDPMs. Our theoretical reflections also indicate that although DDPMs could be suitable for learning a distribution capable of generating plausible user profiles, this behavior is at odds with traditional offline evaluation, which instead rewards models for producing an almost deterministic recommendation list. This is further supported by our observation that the methods we analyze are designed to severely constrain the generative behavior of DDPMs in various ways. While some of the issues we observe may be addressed by altering the architecture, how to achieve this effectively is an open research question that lies beyond the scope of this paper. In sum, our findings indicate that we as a research community still lack the scientific rigor required to ensure real, reliable progress.

The paper is organized as follows. In Section~\ref{sec:previous-studies}, we review previous studies that pointed to various types of methodological issues in recommender systems research. In Section~\ref{sec:methodology}, we explain our research methodology and in Section~\ref{sec:background_DDPM} we provide background information about Denoising Diffusion Probabilistic Models.  The results of our analyses are presented in Section~\ref{sec:results}. The insights and implications of our research are finally discussed in Section~\ref{sec:discussion}.

\section{Previous Studies on Reproducibility and Progress in Recommender Systems}
\label{sec:previous-studies}
Various factors can hinder progress in a field, particularly including a \emph{lack of reproducibility} and \emph{methodological issues}, with a central methodological problem being unsupported assumptions and experimental comparisons involving weak baselines. Several studies in the literature suggest that the field of recommender systems is affected by both of these phenomena.

\paragraph{Reproducibility}
The problem of limited reproducibility of published research on recommender systems algorithms\footnote{In our work, we only focus on algorithms research based on offline experiments. Reproducing results involving humans in the loop is usually considered to be even more challenging.} has been known for many years. For example, already in 2011, Ekstrand et al.~\cite{Ekstrand2011Rethinking} advocated the use of open-source toolkits and presented the LensKit framework to achieve higher levels of reproducibility. In 2013, a dedicated workshop on reproducibility was held at the ACM Conference on Recommender Systems~\cite{Bellogin2013Workshop}. In the context of this  workshop, Konstan and Adomavicius~\cite{Konstan2013Toward} identified several important methodological flaws in the literature of the time, and proposed the use of guidelines and checklists to counter these issues.

Informally speaking, we can say that a work is reproducible if other researchers can obtain (almost) the same results as reported by the authors when repeating the experimental evaluation described in the paper, \idest using the same algorithm, datasets, and evaluation protocol. Given the technical sophistication of many modern recommendation algorithms, accurately re-implementing an algorithm based only on the descriptions in a paper is in most cases challenging and can be error-prone. Furthermore, small variations in the experimental procedures, \eg in terms of data preprocessing and data splitting, may also lead to different evaluation outcomes.\footnote{Similar considerations apply for algorithmic works that do not exclusively focus on accuracy, see \cite{Daniil2024Reproducing} for a recent study in the context of popularity bias mitigation strategies.} Therefore, to ensure reproducibility it is of utmost importance that researchers share the artifacts they used to obtain the results that were reported in a paper.

Unfortunately, it seems that for the majority of published research, the necessary artifacts for reproducibility are not shared, though there are some exceptions. A 2019 study~\cite{ferraridacremaetal2019} showed that only about 39\% of a selection of works published at top-tier venues provided the artifacts (source code and data) allowing researchers to attempt a reproducibility study. A later study published in 2023 on the specific class of recommender systems based on reinforcement learning revealed that only about 21\% of the considered works included the necessary reproducibility artifacts~\cite{Cavenaghi2023RLReproduciblity}. Worryingly, even for papers in which artifacts were shared, some artifacts and important details were often missing, leading to a situation in which only a small fraction of the originally reported numbers could be reproduced.

Notably, reproducibility issues are not specific to recommender systems research. Similar phenomena can be found in the broader machine learning literature~\cite{Semmelrock2024Repro}, and in Artificial Intelligence (AI) in general~\cite{Gunderson2018State}. Gunderson et al.~\cite{Gunderson2018State}, for example, examined as many as 400 papers published in the broad top-level conference series IJCAI and AAAI, and measured different variables which served as indicators for reproducibility, including the documentation of the method, the data, and the experiments. Their analyses, for example, showed that only in very few cases (less than 10\%) the code was shared by the authors. In a later analysis, various \emph{sources of irreproducibility} in machine learning research were identified~\cite{Gunderson2022sources}, and the authors proposed a framework that describes a machine learning methodology to avoid such problems.

Notably, in this work, the term reproducibility is used in a broad interpretation, where reproducibility ultimately means that researchers will arrive at the same conclusions after re-running an experiment.

Overall, as of today, we observe that the level of reproducibility in recommender systems research is not as high as it could be. We are, however, observing positive trends over the years.\footnote{A very recent study focusing on recommender systems based on Graph Neural Networks \cite{TOIS_SIGIR_22} found that in 90\% of the investigated papers, the code was shared at least for the proposed models.}   Sharing code was very rare fifteen years ago, and even less common in times before the existence of popular platforms such as GitHub. Furthermore, reproducibility tracks have become common for conference series such as ACM RecSys or ACM SIGIR, and reproducibility aspects increasingly find their way into the evaluation criteria of review processes. Nonetheless, still more awareness about reproducibility as a cornerstone of scientific research seems required~\cite{bauer2023overcoming}.

\paragraph{Comparisons with Weak Baselines}
In our present work, we benchmark a set of recent diffusion-based models against a number of simpler techniques, strictly following the experimental setup that was used in the original papers. This is important because researchers in the field of recommender systems have many degrees of freedom when designing an experimental setup to demonstrate that the proposed model is more effective than the \emph{state-of-the-art}. This includes the choice of the baselines, the datasets, data preprocessing, and the evaluation metrics used~\cite{Sun2020Arewe,JannachdeSouzaetal2020}. Unfortunately, while most papers proposing new recommendation models claim that they advance the \emph{state-of-the-art}, there is no common understanding of what represents the state-of-the-art. Ultimately, it may become clear that there exists no state-of-the-art in the sense of a small set of models that are better than others across a large set of experimental configurations.

A second, related issue is that of \emph{weak baselines}, that is, comparing a newly proposed method against others that are not effective, either due to model limitations or poor optimization. Note that the term ``weak baseline'' should not be intended as an absolute categorization of a method, but rather should be applied within the context of the specific task and available data. For example, deep learning methods excel with non-structured data while they struggle with tabular data. Then, on non-structured data not all architectures are the same, some are more suited for image processing, others for text or sequence processing, and so on. As another example, a recent reproducibility paper on Graph Neural Networks for collaborative filtering \cite{TOIS_SIGIR_22} observed that, even though graph-based models have a much higher degree of flexibility to accommodate heterogeneous graphs, all the methods analyzed were applied to relatively simple user-item interaction graphs where this flexibility could not play a role. This mismatch between their strengths and the actual data they are applied to did not seem to be recognized or discussed in the analyzed papers. In that scenario, it is entirely possible that methods that are very effective on complex data would be \emph{weak baselines}, because the type of data they handle well is simply not present.

In contrast, ItemKNN was found to be a strong baseline, while on more complex types of data it may be difficult or impossible to apply. This also implies an important rule for experimental design: if a paper claims benefits from complex data, it should include strong baselines that do not rely on it, to validate whether the complex data is helpful at all. Note that the selection of the right baseline models depends on the task, not on the method being proposed. It would be misleading, for example, to select only baselines from the same family as the proposed model, because this leaves open the question of whether that family is a good fit for the task or whether another would be preferable.

One way to mitigate this issue is to rely on standardized benchmarks, which are more common in related fields such as information retrieval. As of today, however, the majority of published research is not based on any of the various evaluation frameworks and toolkits\footnote{Examples of such tools include \href{https://recbole.io/}{RecBole}, \href{https://lenskit.org/}{LensKit}, or \href{https://github.com/sisinflab/elliot} {Elliot}.} that exist today, and which could lead to at least some standardization and comparability of reported results. Recently, efforts by Zhu et al.~\cite{Zhu2022BARS} tried to go even further than evaluation frameworks, by proposing a standardized benchmarking pipeline for reproducible recommender systems research.

So far, such a standardized approach has not found widespread adoption yet, and most published articles seem to be still based on ad-hoc experimental settings. As a result, it becomes difficult to compare experimental results across papers, due to the various experimental configurations that are used by researchers. Indeed, a recent study showed that it was not possible to compare results even between papers published in 2022 and 2023 at the same conference series, ACM SIGIR~\cite{TOIS_SIGIR_22}. Furthermore, it turns out that researchers may often be led by confirmation bias, which leads to experimental design choices that favor their own proposed models. Also, as a potential result of confirmation bias, several studies in the literature suggested that progress in some papers is only achieved because \emph{(i)} the chosen baselines are too weak in general or \emph{(ii)} researchers meticulously tuned their own models but not the baselines. This phenomenon is similar to the observations made by Armstrong et al.~\cite{Armstrong:2009:IDA:1645953.1646031} in the information retrieval community. The phenomenon is also not new in recommender systems, and Konstan and Adomavicius~\cite{Konstan2013Toward} already in 2013 found it not uncommon that ``straw man'' baselines were set up to be beaten by a new model and that hyperparameter settings for the baselines were not reported.

Given these issues, a number of reproducibility studies were published over the past few years, in which recent recommendation models were benchmarked against longer-existing and often simpler baseline models, ensuring that the hyperparameters of \emph{all} compared models were systematically tuned.\footnote{Clearly, an experiment in which this is not done cannot lead to any reliable insights, because, depending on the hyperparameter tuning effort, almost every model can be declared to be a ``winner''~\cite{shehzad2023everyone}.} Two types of such benchmark studies can be identified. One approach is to compare a new model against alternative and well-tuned baselines using the artifacts shared by the authors. This is the approach taken in this paper and in~\cite{ferraridacrema2020tois,DBLP:conf/ijcai/DacremaCJ20} and~\cite{JannachLudewig2017c}. The alternative is to integrate several models into one general evaluation framework and then benchmark these models under identical settings, \eg using the same datasets and evaluation procedures. This approach was followed, for example, in~\cite{anellitop2022,JannachLudewig2017c,ludewiglatifiumuai2020}.

The outcomes of many of these studies are concerning, pointing to persistent problems in our research community. In one of the earlier works from 2017~\cite{JannachLudewig2017c} among these benchmarking papers, it was found that the first version of the first neural model for session-based recommendations, GRU4Rec~\cite{DBLP:journals/corr/HidasiKBT15}, was not more effective than a nearest-neighbor technique. Later studies~\cite{ludewiglatifiumuai2020,latifi2021ins} from 2020 and 2021 then showed that a number of other state-of-the-art neural models of the time for session-based and session-aware recommendation were not consistently more effective than several conceptually simple baseline models. Most recently, it was found that recent session-based models based on Graph Neural Networks often do not reach the effectiveness of well-tuned nearest-neighbor models~\cite{shehzad2024performance}. The study focused on methods proposed for homogeneous graphs with user-item nodes and category information, where nearest-neighbor models can be easily applied and the flexibility of GNNs is likely less important. Applying a new family of models to a dataset that is likely too simple for their strengths to play a role appears to be a rather frequent occurrence. Overall, despite the evidence of their competitive results, they are barely used as baselines in papers today.

For the classical \emph{top-n} recommendation problem, a 2019 study~\cite{ferraridacremaetal2019} reported that, out of seven then-recent neural models, six were less effective than longer-standing matrix factorization models or even simple heuristic baselines. In an extended follow-up study, 11 out of 12 models published in top-tier venues were found to be less effective than properly tuned baselines. One neural model whose effectiveness fell below the baselines was the highly visible Neural Collaborative Filtering (NCF) approach~\cite{He2017Neural}. In~\cite{Rendle2020NeuMF}, Rendle et al.~provided an in-depth comparison of traditional matrix factorization and the NCF framework. Central to their analysis is the traditional use of the dot product to combine two or more embeddings, which is replaced by learned similarities in NCF. Their investigations led to the conclusion that, with proper hyperparameter tuning, using the dot product is favorable over a similarity learning-based approach.

Later on, further research works `revisited' the effectiveness of traditional matrix factorization models. In~\cite{RendleKZK22Revisiting}, Rendle et al. reassessed the effectiveness of \iALS from 2008 for implicit feedback datasets. They found that \iALS, if properly tuned, was still competitive with much more recent models and at the same time benefited from better scalability. Most recently, Milogradskii et al.~\cite{RevisitingBPR2024} re-examined BPR~\cite{DBLP:conf/uai/RendleFGS09}, a popular learning-to-rank method from 2009. Their analyses revealed that BPR, if properly implemented and systematically tuned, can match the effectiveness of different state-of-the-art models of 2024, even though in the literature the results for BPR are often framed as mediocre.

In search of the state-of-the-art for top-n recommendation models, Anelli et al.~\cite{anellitop2022} performed a study in the style of~\cite{ludewiglatifiumuai2020}, in which several models were evaluated in a common framework. Their experiment involved a dozen algorithms from different families and three frequently used datasets. We note that these datasets are relatively small, with the largest one comprising a bit more than half a million interactions.\footnote{After studying models with 100 million interactions in 2006 in the context of the Netflix Prize competition, the community has returned to comparably small datasets due to the limited scalability of many complex neural models.} For these datasets, it turned out that the latest neural models were not among the most effective ones. Instead, the top positions were taken by heuristic neighborhood-based models and comparably simple linear models like \EASER~\cite{DBLP:conf/www/Steck19}, which on average was the most effective in this comparison. In a related study, Anelli et al.~\cite{Anelli2023Myth}~later on focused on a comparison of top-n recommendation models based on GNNs. Their study revealed that the relative effectiveness of the compared models, including traditional ones, varies across datasets. For some datasets, traditional and simpler models work surprisingly well. For others, however, at least some of the more complex GNN models were more effective than traditional ones. Interestingly, the best effectiveness of the GNN models is often achieved by UltraCGN~\cite{Mao2021UltraGCN}, which is much simpler than other GNN-based models. A related study on the use of \emph{heterogeneous} graph neural networks for knowledge-aware recommendation was presented in~\cite{Lv2021ProgressHGN}. In their work, the authors benchmark 12 models based on their original codes. Their study revealed that, differently from the literature, \emph{homogeneous} GNN models exhibit either similar or better effectiveness compared to heterogeneous GNN models when properly tuned.

Overall, our discussions in this section show that there is ample evidence in the literature that algorithm research may actually be stagnating. At least some parts of the reported progress in the literature may be an illusion and the result of methodologically flawed experiments, where well-tuned new models are compared with barely tuned or untuned baselines. The findings of our present analysis of diffusion recommendation models add to this evidence, and show that well-known problems still persist.

\paragraph{Other Methodological Issues}
In addition to the issue of comparisons with weak or untuned baselines, researchers have identified several other methodological issues in the literature. In a recent work, Hidasi and Czapp listed various \emph{pitfalls} and \emph{flaws} in typical offline evaluations~\cite{Hidasi2023Flaws}, using the session-based recommendation task for illustration. The identified issues include using unsuitable datasets for a given task, questionable data preprocessing, data splitting that leads to information leakage, and negative sampling in the test phase.

This latter problem was investigated more deeply in~\cite{Krichene2022Sampled,Dallmann2021Sampling,Canamares2020Sampling}. According to these analyses, the \emph{ranking of models} when comparing their effectiveness can become unreliable when only a subset of the items is considered for ranking in the testing phase. This approach of using ``sampled metrics'' has been around at least since 2008, \eg in~\cite{Koren2008Factorization}. The most worrying aspect of this practice is that it affects the central component of almost any paper on recommendation models, where progress is demonstrated by presenting a ranking of models with the newly proposed model being in the top position.

In an in-depth study on recommender systems evaluation, Sun et al.~\cite{Sun2020Arewe} ask if we are ``evaluating rigorously'' and provide a benchmark setup with standardized procedures for improved reproducibility and reliable progress. An analysis of a larger set of papers published in top conferences confirms that there is no consistency in the literature regarding baselines, datasets, preprocessing steps, data splitting evaluation strategies and so forth, making it difficult to assess the true progress in the field. A concerning side observation of their analysis is that as many as 37\% of the analyzed papers \emph{``directly tune hyper-parameters based upon the performance on the test set.''}
A related observation was made earlier in~\cite{ferraridacrema2020tois}, where the analysis of a dozen neural recommendation models showed that often no information was provided on how the number of training epochs, a central hyperparameter, was determined or how early-stopping was implemented. An investigation of the provided code instead indicates that sometimes the number of epochs is determined during the test phase, and that in a few cases the best accuracy results across epochs are reported in the papers. In terms of hyperparameter tuning, many papers in the literature state that they use one (often small) fixed size for user and item embeddings across models ``for fair comparison.'' However, we argue that this common practice achieves the opposite. The embedding size is a model- and dataset-specific hyperparameter that should be tuned per method rather than fixed across all of them. Different architectures may become most effective at very different embedding dimensionalities, so fixing a single small value can disadvantage some methods more than others and artificially bias comparisons~\cite{shehzad2024performance}. After the optimal embedding dimensionality has been selected, comparing that value across different families of methods can be another direction for discussion, related to the memory and time efficiency of the method. As a concrete example, \citet{RendleKZK22Revisiting} showed that \iALS can achieve very competitive effectiveness with an embedding size in the \emph{thousands}, which is in striking contrast to the commonly used sizes of 64 or 128. The consequence of this result is that using a small embedding size can bias the experimental protocol in a way that disadvantages some matrix factorization methods, making it easier to show that a newly proposed model ``outperforms the state-of-the-art.''

A final problem that may significantly distort the ranking of models when comparing their effectiveness can lie in an incomplete or even incorrect implementation of existing models. Hidasi and Czapp~\cite{HidasiC23ThirdParty} analyzed six `third-party' implementations of the widely used GRU4Rec recommendation model for session-based recommendation. Their investigations revealed that these implementations sometimes deviate heavily from the original architecture and many of them do not implement all features of the original implementation shared by the authors. As a consequence, the obtained results can be dramatically lower than those that one would achieve with the original implementation. Ultimately, such problems can be another contributing factor to the illusion of progress that we observe. Similar observations regarding incomplete and alternative third-party implementations have also recently been made for the popular \MFBPR model~\cite{RevisitingBPR2024}.

In the later sections of our work, we will report if we observed similar methodological issues for more recent works on diffusion recommender models.

\section{Methodology}
\label{sec:methodology}

Our analysis focuses on papers that apply principles of Denoising Diffusion Probabilistic Models (DDPMs) to the domain of recommender systems, in particular to collaborative filtering. We do so because DDPM-based recommenders are a recent and rapidly expanding direction in RecSys with many applications. We also note that DDPMs rely on assumptions suited to tasks such as image generation, while the papers we analyze do not discuss whether collaborative filtering, a very different task, satisfies those assumptions. This study consists of several stages of analysis, summarized in Figure~\ref{fig:method_flow}. First, we identified a set of candidate papers to analyze and verified the availability of the relevant artifacts (source code and datasets) for each of them.
Second, we examined these artifacts to assess whether they were complete, correct, and consistent with the descriptions provided in the papers. Then, we evaluated the \emph{reproducibility} of the results reported in the original papers by re-executing their experiments, following the original experimental protocols and using the provided artifacts. This step is intended to determine whether the reported measurements and conclusions can be independently verified. Finally, we assessed whether the original experimental analysis was conducted in a way that provided sufficient evidence for the claim that the method was more effective than the \emph{state-of-the-art} within the scenario defined by the authors. In particular, we examined whether the proposed models were compared against well-optimized and competitive baselines, and whether best practices were followed in hyperparameter tuning, data preprocessing, and evaluation metrics. By addressing all of these aspects, our study offers a comprehensive analysis of both the reliability and validity of recent diffusion-based recommendation models.

\begin{figure}[t]
\centering
\resizebox{\linewidth}{!}{%
\begin{tikzpicture}[
  >=Latex,
  stepbox/.style={
    rounded corners=3mm, draw, thick, align=left,
    minimum width=3.2cm, minimum height=1.3cm,
    text width=3.2cm, inner sep=6pt, fill=gray!10
  },
  arrow/.style={-{Latex[length=3mm]}, thick}
]
\node[stepbox] (s2) {%
  \textbf{Artifact Analysis}\\
  \begin{itemize}[leftmargin=*, nosep]
    \item Available
    \item Consistent
    \item Runnable
  \end{itemize}
};
\node[stepbox, align=center, above=1.0cm of s2, text width=3.3cm] (s1) {\textbf{Candidate\\ Paper Selection}};

\node[stepbox, right=1cm of s2, text width=3.7cm] (s3) {%
  \textbf{Reproducibility}\\
  \begin{itemize}[leftmargin=*, nosep]
    \item Porting source code
    \item Running experiment
  \end{itemize}
};
\node[stepbox, right=1cm of s3, text width=3.7cm, ] (s4) {%
  \textbf{Result Assessment}\\
  \begin{itemize}[leftmargin=*, nosep]
    \item Consistency
    \item Stability
  \end{itemize}
};
\node[stepbox, right=1cm of s4, text width=3.7cm, ] (s5) {%
  \textbf{Benchmarking}\\
  \begin{itemize}[leftmargin=*, nosep]
    \item State-of-the-art baselines
  \end{itemize}
};

\node (methodlabel) at ($(s2.south)!0.5!(s5.south) + (0,-1.2cm)$) {%
  \begin{minipage}{0.3\linewidth}
    \textbf{Methodology Analysis}
    \begin{itemize}[leftmargin=*, nosep]
      \item Evaluation protocol
      \item Adequacy of the baselines
    \end{itemize}
  \end{minipage}%
};

\begin{scope}[on background layer]
  \node[
    rounded corners=3mm, draw, dashed, thick,
    fill=gray!20, fill opacity=0.25, draw opacity=1,
    inner sep=10pt,
    fit=(s2) (s3) (s4) (s5) (methodlabel)
  ] (methodbox) {};
\end{scope}

\draw[arrow] (s1) -- (s2);
\draw[arrow] (s2) -- (s3);
\draw[arrow] (s3) -- (s4);
\draw[arrow] (s4) -- (s5);

\end{tikzpicture}
}
\caption{Main stages of our analysis workflow.}
\label{fig:method_flow}
\end{figure}

\subsection{Selection of Candidate Papers}
The selection of candidate papers for our reproduction study followed a systematic process. First, we gathered all papers presented in the \emph{Diffusion in RecSys} session at SIGIR '24.\footnote{ACM SIGIR is a top-tier conference with an A* rating in the Australian CORE ranking system, \url{https://sigir-2024.github.io/proceedings.html}}
We then filtered this list to include only papers that (i) proposed a diffusion-based recommender algorithm and (ii) focused on the top-n recommendation problem. We did not include papers utilizing diffusion models for other generative tasks, such as image generation \cite{DBLP:conf/sigir/XuWFMZ024}. Additionally, we included \citet{DBLP:conf/sigir/WangXFL0C23} from SIGIR'23, as it is, to the best of our knowledge, the first diffusion-based recommender model published at SIGIR and is cited by all other candidate papers.
Following this process, we identified four candidate models: DiffRec~\cite{DBLP:conf/sigir/WangXFL0C23}, CF-Diff~\cite{DBLP:conf/sigir/HouPS24},  GiffCF~\cite{DBLP:conf/sigir/ZhuWZX24}, and DDRM~\cite{DBLP:conf/sigir/ZhaoWXSFC24}.

\subsection{Artifact Availability and Consistency}
The next step was to verify whether all candidate papers provide the minimal necessary artifacts, \idest source code and data, required to run the experiments. Specifically, for artifacts to be considered \emph{complete}, they must satisfy the following criteria:
\begin{itemize}
    \item The source code artifacts should include an implementation of both the core model and the experimental pipeline in a \emph{runnable} state. Minor issues, such as missing dependencies or incomplete implementations of the experimental protocol, were corrected when possible.
    \item The data artifacts should contain the original training-test split used in the experiments. The number of items, users, and interactions, as well as the distribution of interactions among items, should be consistent with the preprocessing described in the paper.
    \item If the data artifacts are unavailable, they may still be considered complete if the dataset is publicly accessible and the paper provides sufficient details to accurately replicate the data split.
\end{itemize}

When artifacts were unavailable or we encountered issues using them, we contacted the authors for assistance. We successfully retrieved the artifacts for all candidate papers, though some exhibited inconsistencies.\footnote{In particular, in the original DiffRec source code there was an undefined variable but it was clear how to correct the error. For CF-Diff, the source code included a partial implementation of the guidance which did not allow us to compute the required number of hops. Given that the artifacts included a precomputed guidance, we used that instead (see Section \ref{subsec:CF-Diff}).} If the authors provided us with additional information or different hyperparameter configurations compared to those in the original artifacts, we include them in our analysis.\footnote{In particular, after publishing the first version of this manuscript on arXiv, we were contacted by the authors of DiffRec and DDRM, with whom we had several exchanges to identify the correct configuration to reproduce the results reported in the original papers. While the new configuration they provided improved the results, the overall outcome of our analysis remains unchanged. We report the results with the new configuration in Sections \ref{subsec:DiffRec} and \ref{subsec:DDRM}, and we provide a comparison between the results obtained with hyperparameters derived from the available artifacts and the new ones shared by the authors in Appendices \ref{app:diffrec} and \ref{app:ddrm}.}

\subsection{Reproducibility}
Once \emph{runnable} and \emph{complete} artifacts were retrieved, we incorporated all original implementations into our own framework \cite{ferraridacrema2020tois}, to maintain a consistent evaluation setup. The framework is a Python framework that delivers reproducible, end-to-end experimentation by standardizing data loading, preprocessing and splitting, unified recommendation algorithm interfaces, fast Cython routines, and comprehensive evaluation metrics for accuracy and beyond-accuracy.\footnote{Each metric is computed per user and then averaged over the users who have at least one interaction in the test set.} The original model code was left untouched, while early-stopping and evaluation were handled by our framework.\footnote{Furthermore, to improve efficiency and ensure uniform behavior across models, we substituted the original data sampling with a Cython-based version (\url{https://cython.org/}) that replicates the original sampling logic.} It should be noted that the artifacts of some of the papers we analyze include checkpoints of the model parameters during training. As part of our analysis, we use those checkpoints to verify the correctness of porting the models into our framework, but we do not report their results because our goal is to assess the reproducibility of the whole training process.

Once the porting was completed, the experiments were executed to assess whether the results from the original paper were reproducible.
While reproducibility is generally understood as obtaining numerical results that are \emph{very close} to those reported in the original paper, to the best of our knowledge, there is no established consensus on the exact criteria for determining reproducibility, with different approaches exhibiting different sensitivity \cite{DBLP:journals/ipm/MaistroBSF23}.
Previous studies often deem a paper successfully reproduced if the relative difference between the original results and those obtained from re-running the experiment falls within a fixed threshold. For example, the authors of~\cite{TOIS_SIGIR_22} use a threshold of 2\%. However, diffusion models are generative methods that can exhibit, by construction, a strongly stochastic behavior. In some instances (see GiffCF in Section \ref{subsec:GiffCF}), we observe their effectiveness varying by up to 18\%. With such a high variance, a fixed threshold would be unreliable, and the mean of many measurements bears little meaning without their variance. Notably, all of the original studies report only a single effectiveness metric without any variance estimation, making it difficult to assess the reliability of the reported results.

To assess reproducibility, we evaluated results along two dimensions.
First, we aim to assess whether the results reported in the paper are \emph{statistically compatible} with those we obtain. To this end, we compute the mean $\mu$ and standard deviation $\sigma$ by re-running the experiments 10 times,\footnote{We select the number of training epochs in the first run via early-stopping, and then we use that same value for the remaining 9 runs.} and assess whether the original results reported in the paper fall within one standard deviation of the mean we obtained, \idest
within the range $[\mu - \sigma, \mu + \sigma]$.
This allows us to account for the stochastic behavior of each individual method. However, as the variance increases, it also becomes easier for a method to fall within the interval. For this reason, we also check if the variance exceeds a threshold of 2\% of the mean. If so, we consider the method \emph{unstable}.
This leads to four possible cases (summarized in Figure~\ref{fig:repro_quadrants}):
\begin{itemize}
    \item The results we obtain are \emph{compatible} with those reported in the original paper and the method is \emph{stable}. In this case, the paper can be considered successfully reproduced.
    \item The results we obtain are \emph{compatible} with those reported in the original paper, but the method is \emph{unstable}, exhibiting high variance. While the paper can be considered partially reproducible under our definition, this is an undesirable situation, as the results vary widely between runs, making it difficult to use the method in practice.
    \item The results we obtain are \emph{not compatible} with those reported in the original paper and the method is \emph{stable}. In this case, the paper cannot be considered reproducible. An effectiveness of the method consistently higher or lower compared to what is reported in the original paper suggests that a systematic factor may be influencing the outcomes of the comparison, \eg differences in data splits, hyperparameter configurations, or possible information leakage.
    \item The results we obtain are \emph{not compatible} with those reported in the original paper, and the method is \emph{unstable}, exhibiting high variance. This is the worst possible outcome, as the method is not only not reproducible, but its effectiveness is also highly inconsistent, making it difficult to assess its actual effectiveness.
\end{itemize}

\begin{figure}[t]
\centering
\resizebox{0.7\linewidth}{!}{%
\begin{tikzpicture}[
  box/.style={draw, rounded corners, thick, align=center,
              font=\footnotesize, text width=4.0cm, minimum height=2.1cm},
  brace/.style={decorate, decoration={brace, amplitude=5pt}, thick}
]

\node[box] (q11) at (0,0) {\textbf{Compatible + Stable}\\
\emph{Fully reproducible}.};

\node[box, right=0.6cm of q11] (q12) {\textbf{Not compatible + Stable}\\
\emph{Non reproducible} due to systematic differences (data splits, hyperparameters, leakage).};

\node[box, below=0.6cm of q11] (q21) {\textbf{Compatible + Unstable}\\
\emph{Partially reproducible}. The high variance reduces reliability.};

\node[box, below=0.6cm of q12] (q22) {\textbf{Not compatible + Unstable}\\
\emph{Non reproducible} and highly inconsistent.};

\begin{scope}[on background layer]
  \node[fit=(q12)(q22), rounded corners, inner sep=8pt, fill=badcol2, draw=none] {};
  \node[fit=(q21)(q22), rounded corners, inner sep=8pt, fill=badcol, draw=none] {};
  \node[fit=(q22), rounded corners, inner sep=8pt, fill=badcol2, draw=none] {};
\end{scope}

\draw[brace] ($(q11.north west)+(0,0.55)$) -- node[above=6pt]{\textbf{Compatible}} ($(q11.north east)+(0,0.55)$);
\draw[brace] ($(q12.north west)+(0,0.55)$) -- node[above=6pt]{\textbf{Not compatible}} ($(q12.north east)+(0,0.55)$);
\def\ybraceoffset{0.55} 

\draw[decorate, decoration={brace,amplitude=5pt}, thick]
  ($(q11.south west)+(-\ybraceoffset,0)$) -- ($(q11.north west)+(-\ybraceoffset,0)$);
\node[rotate=90] at ($(q11.west)!0.5!(q11.west)+(-\ybraceoffset-0.55,0)$) {\textbf{Stable}};

\draw[decorate, decoration={brace,,amplitude=5pt}, thick]
  ($(q21.south west)+(-\ybraceoffset,0)$) -- ($(q21.north west)+(-\ybraceoffset,0)$);
\node[rotate=90] at ($(q21.west)!0.5!(q21.west)+(-\ybraceoffset-0.55,0)$) {\textbf{Unstable}};

\end{tikzpicture}
}
\caption{Reproducibility outcomes grouped by compatibility and stability. Highlighted regions indicate undesirable cases.}
\label{fig:repro_quadrants}
\end{figure}

In some cases, only a subset of the reported metrics were reproducible on a given dataset. In such instances, full reproducibility is achieved only if all metrics are successfully reproduced.

It should be noted that all the DDPM models we analyze are evaluated using a slightly unusual setup, which we adopt in our analysis as well to remain consistent with the original papers. As usual, the interaction data is split into three sets: training, validation, and test. A common practice is to optimize the model's hyperparameters on the validation set and then retrain the model on the combined training and validation sets, so that the evaluation on the test set accounts for all available interactions. However, the papers we analyze evaluate models that are only trained on the training set, dismissing the validation data. Another common practice in top-n recommendation scenarios is to remove training interactions from the recommendation lists used for testing, because the user has already interacted with those items in the past and therefore they will not be in the test set (unless repeated interactions are relevant to the task). In the papers we analyze, both the interactions from the training and validation sets are removed from the recommendations during testing. While this practice is not strictly incorrect, it is unclear why interactions from the validation set should be removed given that these interactions were never used for training. A possible explanation is that this was done to simplify the task, since interactions that appear in the validation set cannot also appear in the test set, thereby reducing the number of candidate items to select from.

\subsection{Benchmarking to Assess Improvements over the State-of-the-Art }
While reproducibility focuses on verifying reported results, a related concern is whether the original paper's \emph{claims}, such as being more effective than the state-of-the-art, hold under thorough evaluation. These claims often rely on specific choices of datasets, preprocessing steps, data splits, evaluation metrics, and baselines, which are typically left to the researcher's discretion. This flexibility is understandable, given the wide range of recommendation tasks across domains like e-commerce and streaming services. However, it also introduces the risk of unintentional bias or selective reporting, which underscores the importance of clearly justifying experimental choices and ensuring fair comparisons against strong baselines.
For this reason, this study also aims to assess whether the analysis of the original paper was conducted under state-of-the-art conditions, \idest whether the original paper compared the proposed method against strong baselines. To achieve this, we work within the constraints of the experimental protocol and scenario defined by the original authors, selecting a representative set of strong baselines from different model families and thoroughly optimizing each of them. This comparison is aimed at providing a \emph{lower bound} of what well-known and reasonably computationally efficient methods can already do without relying on any particular domain or scenario-specific adaptation. If the proposed method's effectiveness is considerably worse than the baselines, we can conclude that the analysis in the original paper was \emph{not} conducted in a representative setting and certainly not under state-of-the-art conditions. Consequently, the claim that the method is more effective than prior approaches is unsubstantiated.
It is important to stress that this does not imply that the method can \emph{never} be competitive under any conditions (indeed, it would be impossible to empirically prove such a negative statement) but only that the experimental analysis in the original paper did not provide enough evidence that it could.

If the analysis in the original paper was not conducted under state-of-the-art conditions, it is likely that none of the methods, including the proposed one, were thoroughly optimized. Had the baselines been more effective, it is reasonable to assume that the authors would have invested additional effort in further fine-tuning their proposed method. As a result, the experimental findings cannot be considered reliable, and the reported results for all methods are likely suboptimal. Similarly, the hyperparameters identified as optimal for the proposed method are unlikely to be truly optimal and should always be used with this limitation in mind.

Determining whether a method could be more effective than stronger baselines with proper optimization would require re-running hyperparameter tuning and training with an adequate search space, which will depend on the model's characteristics, falling outside the scope of this study. Such an investigation would shift the focus toward a \emph{generalizability} study and, given the high computational cost of most modern complex methods, would impose an unsustainable burden on researchers conducting studies of this nature.
Ultimately, we argue that the \emph{burden of proof} lies with the authors proposing a new method to demonstrate that it is truly more effective than a reasonable state-of-the-art under the experimental protocol and scenario they have identified.

\subsubsection{Baselines}
In our benchmark experiments, we considered a total of 18 baseline models from different families, selected based on earlier and related reproducibility studies \cite{TOIS_SIGIR_22,ferraridacrema2020tois,anellitop2022}. Note that we chose these baselines because they are very effective for the top-n collaborative filtering task we focus on. For tasks with more complex data, an expanded or different set of baselines may be more appropriate. To keep the presentation of experimental results focused, we describe and report results for only a subset of 9 baselines in the main text; full descriptions and results for all 18 baselines are provided in the Appendix \ref{app:complete-results}.
Below, we briefly describe the 9 baselines included in the main text:
\begin{itemize}[left = 0pt]
    \item \textbf{TopPop}: A non-personalized method recommending to all users the most popular items the user has not yet interacted with.
    \item \textbf{UserKNN}: A user-based nearest-neighbor algorithm~\cite{DBLP:conf/cscw/ResnickISBR94}, using cosine similarity and shrinkage~\cite{bell2007improved}.
    \item \textbf{ItemKNN}: An item-based nearest-neighbor algorithm~\cite{DBLP:conf/www/SarwarKKR01}, using cosine similarity and shrinkage~\cite{bell2007improved}.
    \item \textbf{\pbeta}: A graph-based method that uses a two-step random walk from users to items and vice versa, where transition probabilities are computed from the normalized ratings~\cite{DBLP:journals/tiis/PaudelCNB17}.
    \item \textbf{GF-CF}: A graph-based method that is based on a low-pass filter and has a closed form solution~\cite{DBLP:conf/cikm/ShenWZSZLL21}.
    \item \textbf{\EASER}: An ``embarrassingly shallow'' linear model with strong connections with autoencoders and a closed form solution~\cite{DBLP:conf/www/Steck19}.
    \item \textbf{SLIM}: An item-based model that uses linear regression to compute item similarities~\cite{DBLP:conf/icdm/NingK11}.\footnote{In our implementation, we optimize the ElasticNet loss.}
    \item \textbf{\MFBPR}: A matrix factorization method based on the \emph{Bayesian Personalized Ranking} (BPR) loss~\cite{DBLP:conf/uai/RendleFGS09}.
    \item \textbf{\iALS}: A matrix factorization method for ranking tasks based on alternating least-squares \cite{DBLP:conf/icdm/HuKV08}.
    \item \textbf{MultVAE}: A variational autoencoder that assumes a multinomial likelihood for user-item interactions \cite{DBLP:conf/www/LiangKHJ18}.
\end{itemize}

Note that occasionally the results for \textbf{\EASER} may be missing due to its memory requirements exceeding the 64 GB available on our server (see Section \ref{subsec:computational_resources}).

\subsubsection{Hyperparameter Optimization} To ensure a fair and rigorous evaluation, we performed hyperparameter optimization for all baseline methods using Bayesian optimization, as implemented in Scikit-Optimize. We explored a total of 50 configurations, with the first 16 chosen randomly as initial points. Each configuration was allowed a maximum training time of 3 days. If this limit was exceeded, the training was stopped, the model evaluated and the process moved on to the next configuration. The optimization had a maximum time budget of 14 days, and was truncated early if it reached this threshold. The trained model that led to the best effectiveness on the validation data, according to an evaluation metric and given cut-off that depends on the paper, was then evaluated on the test data. For consistency with the original papers, we adopt the same evaluation protocol: models are trained only on the training split and \emph{not} retrained on the union of training and validation; during validation we filter items seen in the training interactions, and during testing we filter all items seen in either the training or the validation interactions. The search spaces for all methods follow the ranges established by \citet{ferraridacrema2020tois} (see Appendix \ref{app:baseline-hparams} for details).

\subsubsection{Early-stopping} For baseline methods requiring iterative optimization, we employed early-stopping to determine the optimal number of training epochs. Specifically, model effectiveness was evaluated on the validation set every five epochs, and training was halted if no improvement was observed for five consecutive evaluations. The optimal number of epochs was chosen as the one that led to the most effective model. This approach prevents overfitting while ensuring fair comparisons across models.

\subsubsection{Computational Resources} \label{subsec:computational_resources}
All DDPM-based models as well as the MultVAE baseline were run on a GeForce RTX 3090 GPU with 24 GB of RAM, while all remaining baselines were run on an i9-13900K CPU with 64 GB of RAM.

\section{Background on Denoising Diffusion Probabilistic Models}
\label{sec:background_DDPM}
The aim of this section is to provide the necessary background to understand the general idea of Denoising Diffusion Probabilistic Models (DDPMs), the assumptions behind them, the reasons for their popularity, and how they have been used to build the recommendation models proposed in the papers we analyze. Understanding these principles is also important because, as we will later discuss, the way DDPMs are applied in these papers raises conceptual concerns about their suitability for the top-n recommendation task.

Diffusion models are a class of generative models that learn to reverse a gradual noising process applied to data, effectively generating new samples from random noise. Among these, DDPMs have gained prominence due to their simplicity and strong empirical performance, particularly in domains such as image and video generation. In the context of this work, we focus on DDPMs, as they form the basis of the methods used in the recommendation papers we analyze.
DDPMs \cite{DBLP:conf/icml/Sohl-DicksteinW15,Ho2020Diffusion} were originally developed to address a central challenge in machine learning and statistics: modeling a complex data distribution while preserving tractable sampling (\idest generating new samples) and probability evaluation. In many applications, especially when dealing with high-dimensional data like images or videos, the data distribution is modeled using highly complex functions (\idest deep neural networks) that do not lend themselves easily to direct sampling or statistical inference.

The key idea behind DDPMs, inspired by non-equilibrium statistical physics, is to learn a transformation that maps a simple, tractable distribution (such as a Gaussian) to the complex data distribution. For example, a diffusion model for image generation might learn a mapping between a standard normal distribution and a distribution over images, so that new images can be generated simply by sampling from the Gaussian and then applying the learned transformation.

In the following, the structure of DDPMs, their training and sampling procedures are described, along with their application in recommender systems.

\subsection{Structure}
\label{subsec:structure}
Diffusion models achieve the objective of approximating complex distributions through a two-steps procedure:
\begin{enumerate}
    \item Data $X^{(0)}$ coming from the original complex distribution are gradually corrupted by a stochastic forward process $X^{(0)}, X^{(1)}, \dots, X^{(T)}$. This is equivalent to incrementally adding a small amount of noise at each step until the original distribution is effectively “destroyed.” This process gradually transforms the original data distribution into a known, tractable distribution as $t$ increases. By the final step $T$, the data have effectively lost all recognizable structure.
    \item A learnable backward (reverse) stochastic process $X^{(T)}, \hat{X}^{(T-1)}, \dots, \hat{X}^{(0)}$ then reverses this corruption, reconstructing the original data distribution (\idest $\hat{X}^{(0)} \approx X^{(0)}$) from the tractable $X^{(T)}$.
\end{enumerate}

Both the forward and backward processes are modeled as Markov chains, characterized by transition kernels. The forward process kernel is defined as:
\begin{equation*}
q\left(x^{(t)} \bigm\vert x^{(t-1)}; \beta_{t}\right),
\end{equation*}
for $t=1, \dots, T$, where $\beta_{t}$ is the diffusion rate controlling the amount of noise added at each step. Analogously, the backward (or reverse) process kernel is defined as:
\begin{equation*}
p_{\theta}\left(\hat{x}^{(t-1)} \mid \hat{x}^{(t)}; \theta\right),
\end{equation*}
for $t=T,\dots,1$. In practice, the forward process is often fixed, while the reverse process is parameterized by learnable functions (often neural networks).

\paragraph{Gaussian Diffusion Models}
Diffusion models are most commonly implemented using Gaussian kernels \cite{OntheTheoryofStochasticProcesses:112810}. The forward process kernel is typically given by:
\begin{equation*}
q\left(x^{(t)} \bigm\vert x^{(t-1)}; \beta_t\right)=\mathcal{N}\Bigl(x^{(t)};\sqrt{1-\beta_t}\,x^{(t-1)},\,\beta_{t}\mathrm{I}\Bigr),
\end{equation*}
which is equivalent to gradually adding Gaussian noise to the data. The diffusion rate $\beta_t$ regulates the noise injected at each step. After a sufficient number $T$ of steps, the forward process yields a distribution $q\left(x^{(T)}\right)$ that is approximately a standard Gaussian, \idest
\begin{equation*}
q\left(x^{(T)}\right)\approx\mathcal{N}\Bigl(x^{(T)};0,\mathrm{I}\Bigr).
\end{equation*}

The reverse process is parameterized by a similar Gaussian kernel:
\begin{equation*}
p_{\theta}\left(\hat{x}^{(t-1)} \bigm\vert \hat{x}^{(t)}; \theta\right)=\mathcal{N}\Bigl(\hat{x}^{(t-1)};\mu_{\theta}(\hat{x}^{(t)},t),\,\Sigma_{\theta}(\hat{x}^{(t)},t)\Bigr),
\end{equation*}
where the mean $\mu_{\theta}(\cdot)$ and covariance $\Sigma_{\theta}(\cdot)$ are learned. A common design choice is to fix $\Sigma_{\theta}(\hat{x}^{(t)},t)=\beta_{t}\mathrm{I}$ for $t=T,\dots,1$ to reduce the number of learnable parameters.

\paragraph{Conditional Diffusion Models}
In their classical formulation, diffusion models generate samples from the learned data distribution. For example, a diffusion model trained on pictures of real-world landscapes will generate \emph{any} plausible landscape. However, many applications (such as image or video generation) require controlled synthesis, \idest we might want to specify whether the landscape should contain mountains, trees, or any other feature. To this end, a mechanism known as \emph{guidance} is introduced.

Because diffusion models are highly flexible, they can be extended to model conditional distributions $p_\theta(x \mid y)$, where $y$ represents any auxiliary information (\idest text descriptions, other images, or contextual metadata). This additional guidance directs the reverse process toward outputs that are consistent with the conditioning signal.

Mathematically, the formulation for conditional diffusion models closely mirrors that of the unconditional case. The primary modification is in the reverse process, where the learnable functions are augmented to accept $y$ as an input. Formally, the reverse kernel becomes:
\begin{equation*}
p_{\theta}\left(\hat{x}^{(t-1)} \mid \hat{x}^{(t)}, y\right)=\mathcal{N}\Bigl(\hat{x}^{(t-1)};\mu_{\theta}(\hat{x}^{(t)}, y, t),\,\Sigma_{\theta}(\hat{x}^{(t)}, y, t)\Bigr).
\end{equation*}
As before, a popular choice is to fix $\;\Sigma_{\theta}(\hat{x}^{(t)}, y, t)=\beta_{t}\mathrm{I}\;$ for simplicity.

This conditional probability modeling enables the model to generate samples conditioned on $y$ by guiding the reverse process accordingly.

\subsection{Training}
\label{subsec:training}
Training a DDPM usually consists of gradually mapping clean data \(X^{(0)}\) to a simple, tractable distribution at time \(T\), using the forward noising process, which is usually fixed and not learned. The reverse process, parameterized by a neural network, that denoises step by step from \(\hat{X}^{(t)}\) to \(\hat{X}^{(t-1)}\), is then learned. In practice, the model is optimized to learn a reverse step that matches the reverse of the forward step. The key point is simple: forward is fixed, backward is learned.

\paragraph{Loss}
The training loss follows from the goal of matching the final model distribution \(p_{\theta}(\hat{x}^{(0)})\) to the original data distribution \(q(x^{(0)})\). Discussing it in detail is beyond the scope of the paper. In this section we will however provide an intuition for its goal. A standard choice is to minimize their Kullback-Leibler (KL) divergence, which is equivalent to maximizing the expected log--likelihood of the backward process on the data \(\mathbb{E}_{x\sim q(x^{(0)})}[\log p_{\theta}(x)]\). This is mathematically equivalent to optimizing the evidence lower bound (ELBO), which can be expressed in tractable terms, in particular as a sum of KL divergences for each time step between the ideal reverse step and the learned one, plus some constants.
In the Gaussian case this becomes especially simple, making the approximation fully tractable.

\paragraph{Convergence}
\label{par:convergence}
It is important to note that the approximation capabilities of DDPMs have strong theoretical foundations. First of all, with the limit of an infinite number of diffusion steps, the original data distribution $X^{(0)}$ is totally corrupted to pure noise, since the data become more and more ``blurred'', and any structure disappears. This can be easily shown as $T\to\infty$ for the last forward diffusion step. Let $\epsilon\sim\mathcal{N}(0, I)$, and $0 < \beta_t < 1\ \forall t$. Then:
$$
\lim_{T \to \infty}X^{(T)} = \lim_{T \to \infty}\sqrt{1-\beta_T}\; X^{(T-1)} + \sqrt{\beta_T}\; \epsilon=\lim_{T \to \infty}\sqrt{\prod_t^T(1-\beta_t)}\;X^{(0)} + \sqrt{1-\prod_t^T(1-\beta_t)}\; \epsilon=\epsilon
$$
for each sequence $\{\beta_t\}_{t=1}^T$ that satisfies $\lim_{T \to \infty}\prod_t^T(1-\beta_t) = 0$, \eg monotonically increasing sequences. Moreover, in the limit of infinite diffusion steps and infinitesimal diffusion rate $\beta_t$ \cite{OntheTheoryofStochasticProcesses:112810}, the two-stage Markov chain structure ensures that the distribution of each latent random variable $\hat{X}^{(t)}$ in the forward chain has the same distribution as $\hat{X}^{(t)}$ in the backward process, ensuring the convergence of the backward process to the true data distribution. In practice, the number of diffusion steps is finite. A good approximation is realized with a large number of diffusion steps, so that ``\emph{the longer the trajectory the smaller the diffusion rate $\beta$ can be made}'', as reported in \citet{DBLP:conf/icml/Sohl-DicksteinW15}.

\subsection{Sampling}
\label{subsec:sampling}
Since DDPMs are generative models, inference corresponds to sampling. The forward process is discarded, \(X^{(T)}\) is sampled from the simple distribution at the end of the forward process (\eg a standard Gaussian), and the learned reverse process is run down to \(\hat{X}^{(0)}\). Each step removes a bit of noise according to the schedule, so the final sample \(\hat{X}^{(0)}\) follows the complex data distribution learned during training. This can be done with many small steps.

Note that sampling is unconditional by default. There is no input data at sampling time: \(X^{(T)}\) is drawn from the simple distribution (\eg \(\mathcal{N}(0,I)\) in the case of Gaussian DDPM) and the backward reverse process is run to obtain \(\hat{X}^{(0)}\). When using guidance, an auxiliary signal \(y\) is supplied to the backward process kernels so that the final samples follow \(p_{\theta}(\hat{x}^{(0)}\mid y)\). The sampling procedure itself is the same, only the reverse denoiser is now conditional.

\subsection{Diffusion Models in Recommender Systems}
\label{sec:background_DDPM_recsys}
In recommender systems, the primary objective is to model complex distributions over user preferences or interactions in order to predict or generate effective recommendations. Diffusion models offer a promising framework, as they can represent complex, high-dimensional distributions in a tractable manner.

The common approach adopted by the papers we analyze is to apply the forward diffusion process to user profiles or their latent representations. For instance, one can corrupt a user's interaction history by gradually adding noise and then learn a reverse process that recovers the underlying preference signal. In this setup, the model learns the distribution $q(x^{(0)})$ of user profiles and can be used to generate or complete profiles from noisy inputs \cite{DBLP:conf/sigir/WangXFL0C23, DBLP:conf/sigir/HouPS24, DBLP:conf/sigir/ZhuWZX24}, instead of pure noise.
Another approach is to perform diffusion in the latent space of user and item embeddings. Here, different diffusion processes may be designed for users and items, with the goal of capturing complex collaborative and content-based signals \cite{DBLP:conf/sigir/ZhaoWXSFC24}.

In principle, the  guidance signal $y$ in conditional diffusion can incorporate a range of useful heterogeneous information, such as collaborative filtering signals, contextual factors (\eg time or location), or item content. By conditioning the reverse diffusion process on $y$, the model can generate recommendations that reflect both the user's past behavior and additional contextual constraints.
In fact, all methods that we analyze in this work, except for the method by~\citet{DBLP:conf/sigir/WangXFL0C23}, rely on such a guidance mechanism. For instance, \citet{DBLP:conf/sigir/HouPS24} treat the user-item ratings matrix as a bipartite graph (with users and items as separate node types) and define guidance based on the probability of reaching each node within a chosen number of graph hops. \citet{DBLP:conf/sigir/ZhuWZX24} instead provide the user profile itself as guidance, stochastically dropping a percentage of interactions to introduce controlled noise. Finally, \citet{DBLP:conf/sigir/ZhaoWXSFC24} perform two separate diffusion processes: one for user embeddings and one for item embeddings, using each denoised embedding as guidance for the other diffusion process.

However, as discussed, diffusion models rely on several specific assumptions, which makes their application to new domains nontrivial, including recommendation systems with their unique data characteristics and semantics. As we will discuss in detail in Section~\ref{subsec:DDMP-critical}, the adaptation of DDPMs for recommendation task in the papers analyzed in this work seems surprising and raises several fundamental concerns, including \emph{(i)} a mismatch between the generative goals of DDPMs and the nature of offline evaluation in recommender systems, \emph{(ii)} the use of overly constrained guidance signals that reduce generative flexibility, and \emph{(iii)} questionable assumptions about the type of noise present in user interaction data.

%
\section{Results}
\label{sec:results}
In this section, we present the results of our analysis of the selected papers along different dimensions. We follow a structured approach in which we discuss the following aspects for each paper:

\begin{itemize}
    \item \emph{Datasets}: We report which datasets were used in the paper, which preprocessing and splitting strategies were applied, and if there is any justification for the selection of the datasets.
    \item \emph{Artifact Availability and Consistency}: Here, we evaluate the availability and consistency of the shared artifacts, including source code, data, and model checkpoints.
    \item \emph{Methodological Issues}: In this section, we investigate if there are any potential methodological issues in the experimental setup, \eg in terms of hyperparameter tuning. 
    \item \emph{Reproducibility}: We analyze the reproducibility of the results reported in the original papers, assessed through multiple re-runs using the  original artifacts and experimental protocol.
    \item \emph{Benchmarking Results}:  Here, we report the outcomes of a comparative evaluation of the proposed model against well-tuned baseline models.
\end{itemize}

Overall, we conducted an extensive set of experiments. We report the results of approximately 580 trained models, which correspond to a total of 10,282 models when accounting for hyperparameter optimization. The total computational time sums up to 1.87 years. In this section, we highlight representative experimental results for each examined paper to keep our presentation concise and focused. The complete results 
are provided in Appendix \ref{app:complete-results}, and we publicly share the source code used for all experiments on GitHub.\footnote{\url{https://github.com/remaplab/TORS26_Reproducibility-DDPMs}}

\subsection{Diffusion Recommender Model -- DiffRec (SIGIR '23)}
\label{subsec:DiffRec}
\citet{DBLP:conf/sigir/WangXFL0C23} propose DiffRec, a collaborative filtering method that applies standard Gaussian diffusion (without guidance) to user profiles. Three additional variants of the model are presented: L-DiffRec, which partitions user profiles into segments based on item clusters, where item embeddings from a pretrained model are used for clustering, and applies diffusion to each segment after projecting it into a latent space; T-DiffRec, which applies diffusion to user profiles that are re-weighted using temporal information; and LT-DiffRec, which combines L-DiffRec and T-DiffRec.

\paragraph{Datasets}
DiffRec was evaluated on three publicly available datasets consisting of user-item ratings: MovieLens-1M,\footnote{\url{https://grouplens.org/datasets/movielens/1m/}.} Yelp,\footnote{\url{https://www.yelp.com/dataset/}.} and Amazon-Books.\footnote{\url{https://jmcauley.ucsd.edu/data/amazon/}.} No particular justification for the selection of these datasets was provided in the paper. In a preprocessing step, all interactions with a rating value lower than 4 were removed from the datasets.
The remaining interactions were converted to ones. These were then sorted chronologically and split into training, validation, and test sets. The split ratio reported in the original paper is 7:1:2.\footnote{An analysis of the shared datasets however rather suggests a 7:2:1 ratio.} Besides this ``clean'' version of the datasets, two alternative ``noisy'' versions of each dataset were created from the clean dataset. The first version is called ``natural noise'' and was created by adding interactions with rating values below 4 to the training and validation sets. The second version is called ``random noise'' and was created by adding random interactions to the training and validation sets. These noisy training set versions were created to study the effectiveness of the model in the presence of noise. The test sets remain unchanged.

We note that DiffRec and L-DiffRec were evaluated on all clean and natural noise dataset versions, while the time-aware variants, T-DiffRec and LT-DiffRec, were evaluated only on the clean dataset versions. Only DiffRec was also evaluated on the random noise dataset versions. In total, there should be 21 different experimental configurations, but the results of four of them are omitted from the original paper for space reasons, ending up with 17 experimental configurations to reproduce. The pretrained item embeddings required for the clustering done in L-DiffRec and LT-DiffRec are obtained using LightGCN~\cite{DBLP:conf/sigir/0001DWLZ020}.

\paragraph{Artifacts Availability and Consistency}
The authors share a link to a public GitHub repository.\footnote{\url{https://github.com/YiyanXu/DiffRec}} The repository contains both the implementation of the proposed models as well as the models checkpoints used in the experiments. Data splits are shared for the clean and ``natural noise'' datasets. Data splits for the ``random noise'' datasets are missing. Since the code for splitting and preprocessing the data is not provided, we could not conduct the reproducibility experiments on the random noise dataset versions. Additionally, the repository does not include implementations of the baseline models or the code used for training and optimizing them.

Regarding consistency, we observed discrepancies between the dataset split ratios reported in the paper and those found in the shared splits. The paper states a 7:1:2 ratio, but our analysis suggests the actual split is closer to 7:2:1. Moreover, we discovered that the Yelp natural noise dataset contains a small number of overlapping interactions between the training, validation, and test sets (5,192 interactions, less than 0.004\%). Based on their small number we expect minimal impact on the model's effectiveness. Nonetheless we preserved the test set as-is and eliminated from the validation set the interactions that were present in the test set, followed by removing from the training set the interactions that were already present in the validation and test sets.
Another issue concerns the pretrained item embeddings required for L-DiffRec and LT-DiffRec. While these embeddings are shared, no details are provided about how they were generated, including which LightGCN implementation was used and how hyperparameters were chosen. For this reason, it is not possible to conduct a full reproducibility attempt for this part.\footnote{After publishing the first version of this manuscript on arXiv, we were contacted by the authors, with whom we had several exchanges to identify the correct configuration to reproduce the results reported in the original papers. While the new configuration they provided improved the results, the overall outcome of our analysis remains unchanged. In this section, we report the results obtained with the new configuration, and we provide a comparison between this configuration and the hyperparameters we originally derived from the available artifacts in Appendix \ref{app:diffrec}.}

Overall, we find that the provided artifacts are \emph{not complete} and \emph{not fully consistent} with the descriptions in the paper.

\paragraph{Methodological Issues}
The paper proposing DiffRec reports seven baselines, which include six neural models for top-n recommendation, one neural model for sequential recommendation and an earlier model based on matrix factorization (\MFBPR)~\cite{DBLP:conf/uai/RendleFGS09}. However, not all of them are used for all experimental configurations. The paper reports the ranges for some of the hyperparameters for almost all models; however, those ranges are often restricted to three or four values and some fundamental ones \eg embedding sizes, are  not consistently reported for all baselines.
For other hyperparameters, fixed (default) values are used without sufficient justification and the paper does not mention how the optimal hyperparameters for the baselines were determined.

Due to this incomplete reporting, it remains unclear if sufficient efforts were put into tuning the baseline models. Therefore, the danger exists that the proposed model was benchmarked against weakly-tuned baselines, which we consider a potential methodological issue. In terms of the selection of the baselines, the authors mention that their set includes both generative and non-generative models. A further discussion regarding the choice of the baseline models is not provided.

\begin{table}[ht]
\centering
\footnotesize
\begin{tabular}{ll|cc|cc}
\toprule
& \textbf{Dataset} & \multicolumn{2}{c}{\textbf{Compatible}} & \multicolumn{2}{c}{\textbf{Not Compatible}} \\
& & \textbf{Stable} & \textbf{Unstable} & \textbf{Stable} & \textbf{Unstable} \\
\hline
\multirow[t]{6}{*}{DiffRec} 
  & MovieLens-1M (clean) & 0/4 & 0/4 & 4/4 & 0/4  \\
  & Yelp (clean)         & 4/4 & 0/4 & 0/4 & 0/4  \\
  & Amazon-Books (clean) & 4/4 & 0/4 & 0/4 & 0/4  \\
  & MovieLens-1M (natural noise) & 0/4 & 0/4 & 2/4 & 2/4  \\
  & Yelp (natural noise)         & 1/4 & 0/4 & 3/4 & 0/4  \\
  & Amazon-Books (natural noise) & 4/4 & 0/4 & 0/4 & 0/4  \\
\hline
\multirow[t]{6}{*}{L-DiffRec}
  & MovieLens-1M (clean) & 0/4 & 0/4 & 3/4 & 1/4  \\
  & Yelp (clean)         & 0/4 & 0/4 & 4/4 & 0/4  \\
  & Amazon-Books (clean) & 0/4 & 1/4 & 2/4 & 1/4  \\
  & MovieLens-1M (natural noise) & 1/4 & 3/4 & 0/4 & 0/4  \\
  & Yelp (natural noise)         & 4/4 & 0/4 & 0/4 & 0/4  \\
  & Amazon-Books (natural noise) & 0/4 & 0/4 & 0/4 & 4/4  \\
\hline
\multirow[t]{3}{*}{T-DiffRec}
  & Yelp (clean)         & 2/4 & 0/4 & 2/4 & 0/4  \\
  & Amazon-Books (clean) & 4/4 & 0/4 & 0/4 & 0/4  \\
\hline
\multirow[t]{3}{*}{LT-DiffRec}
  & Yelp (clean)         & 1/4 & 0/4 & 3/4 & 0/4  \\
  & Amazon-Books (clean) & 1/4 & 3/4 & 0/4 & 0/4  \\
\hline
\textbf{Total} & & \textbf{26/64} & \textbf{7/64} & \textbf{23/64} & \textbf{8/64}  \\
\bottomrule
\end{tabular}
\caption{Reproducibility outcomes of DiffRec and its variants.}
\label{tab:summary-diffrec}
\end{table}

\paragraph{Reproducibility}
We could successfully execute the provided source code for the proposed models on the shared datasets.
We recall that there are four DiffRec variants, three datasets, and three variations of each dataset, but not all DiffRec variants are evaluated on all variations of each dataset. This corresponds to a total of 17 different experimental configurations, and for each of them, four metric values are considered: Recall and NDCG at cut-off lengths of 10 and 20. Analyzing the outcomes of several executions of each configuration, we find that only a subset of the results reported in the original paper are reproducible according to our definition, as detailed in Table \ref{tab:summary-diffrec}. In summary:

\begin{itemize}
    \item Since the random noise data splits were not shared (some results are even not reported in the main paper), for three configurations we could not run the experiment at all.
    \item In the case of the basic DiffRec method, we could reproduce the results for three out of six datasets: Yelp clean (NDCG@10 of $0.0364 \pm 0.0001$ with the original paper reporting $0.0363$), Amazon-Books clean (see Table \ref{tab:DiffRec-amazon-book_clean_original-result}), and Amazon-Books natural noise (NDCG@10 of $0.0333 \pm 0.0004$ with the original paper reporting $0.0335$). Our results on Yelp natural noise show a higher effectiveness, but are \emph{not compatible} with those reported in the original paper (NDCG@10 of $0.0313 \pm 0.0001$ with the original paper reporting $0.0309$), except for Recall@20, which is the only \emph{reproducible} measurement. The same is true for MovieLens-1M clean, where the large absolute difference between the results we obtain and those reported in the paper indicates that the results are \emph{not compatible} (MovieLens-1M clean has an NDCG@10 of $0.0856 \pm 0.0015$ with the original paper reporting $0.0901$).  It should be noted that the variance of DiffRec tends to be very tight, with few exceptions on the MovieLens-1M natural noise dataset. Overall, the results on MovieLens-1M clean and natural noise, as well as on Yelp natural noise are \emph{non reproducible} mostly due to a systematic shift.
    \item For L-DiffRec, out of six experiments the only one that produced \emph{reproducible} results was for the Yelp natural noise dataset (NDCG@10 of $0.0313 \pm 0.0003$ with the original paper reporting $0.0311$). All the results on the MovieLens-1M clean, Yelp clean and Amazon-Books natural noise datasets were \emph{not compatible} with those reported in the original paper, as the metric values reported in the paper were outside the tolerance interval we computed based on our experiments; some of these metrics were also \emph{unstable}, \eg all metrics of Amazon-Books natural noise. The results on Amazon-Books clean were mostly \emph{not compatible}, with only one exception. Finally, the results on MovieLens-1M natural noise were all \emph{compatible} but mostly \emph{unstable} and therefore we consider them as being \emph{partially reproducible}. In most cases, the mean of the results we obtained was lower by 3-5\% compared to what is reported in the paper, with the exception of MovieLens-1M natural noise, where we obtained better effectiveness by more than 2\%. 
    \item For T-DiffRec, we obtained more positive results. For the Amazon-Books clean (see Table \ref{tab:DiffRec-amazon-book_clean_original-result}) dataset the results are \emph{compatible} and \emph{stable}, \idest within the interval we defined based on our experiments and the variance itself is very low, therefore we consider these results \emph{fully reproducible}. For the Yelp clean dataset instead, while all results are \emph{stable}, only half of them are also \emph{compatible} with what is reported in the paper, therefore we consider them as \emph{partially reproducible}.
    \item For LT-DiffRec, we obtained slightly more negative results. In the case of Amazon-Books clean (see Table \ref{tab:DiffRec-amazon-book_clean_original-result}), we found that the effectiveness is very close to that reported in the original paper. However, the variance is relatively large (over 2.5\% of the mean) on three metrics out of four, indicating the method is rather \emph{unstable}. For Yelp clean, the results reported in the paper are very close in absolute value and exhibit small variance; however, three out of four metrics are slightly outside the compatibility interval we defined in Section \ref{sec:methodology} (NDCG@10 of $0.0361 \pm 0.0002$ with the original paper reporting $0.0369$). Overall, the results are \emph{partially reproducible}.

\end{itemize}

In total, we could fully reproduce five and partially reproduce three out of 17 configurations based on the shared artifacts.
In Table \ref{tab:DiffRec-amazon-book_clean_original-result}, we provide the results for the clean Amazon-Books dataset as an example, since experiments for this dataset exhibited the lowest reproducibility among the datasets that we tested with the DiffRec variants. A complete set of experimental results is provided in Appendix \ref{app:diffrec}. Notably, DiffRec and its variants exhibited varying levels of variance, significant in some cases and very low in others, as shown in Table~\ref{tab:DiffRec-amazon-book_clean_original-result}. In this case, the narrow tolerance intervals have the effect that the obtained results are classified as not compatible, despite the absolute values being relatively close to those reported in the paper.

\tableArticleResult{DiffRec}{DiffRec}{amazon-book_clean_original}{Amazon-Books (clean)}{1}{}{0}

\paragraph{Benchmarking Results}
Table~\ref{tab:DiffRec-amazon-book_clean_original-result} also shows the effectiveness results that we obtained for our own set of baseline models, which we systematically tuned and evaluated on all datasets. Notably, we find that all of the conceptually very simple neighborhood and graph-based models, shown in the top rows of the table, consistently outperform DiffRec and all its variants on this dataset. Interestingly, the very old ItemKNN model is almost consistently leading to the best results on the Amazon-Books clean dataset.

Looking at the effectiveness of our other baseline models, we find that SLIM also performs very well, whereas the matrix factorization models \MFBPR and \iALS lead to very low effectiveness in this dataset. This latter aspect was also observed and reported by the authors of DiffRec. For other datasets like MovieLens-1M, \MFBPR and in particular \iALS lead to more competitive results. Finally, we note that the results for the \EASER model are missing from the table, because the significant memory requirements of the model exceeded the 64 GB available on our hardware.

The effectiveness results for DiffRec and its variations on the other datasets, as found in Appendix \ref{app:diffrec}, can be summarized as follows. For the natural noise version of the Amazon-Books dataset, the same observations were made as for the clean version in Table~\ref{tab:DiffRec-amazon-book_clean_original-result}, \idest all simple models are more effective than DiffRec and its variant L-DiffRec. For the Yelp clean and natural noise datasets, the effectiveness of DiffRec and its variants is at the level of ItemKNN, a method that is at least 25 years old, but it is outperformed by various models like \pbeta, \iALS or GF-CF. For the clean MovieLens-1M dataset, the effectiveness results reported in the original paper for DiffRec would be competitive with many models, but, as mentioned, DiffRec and the variant L-DiffRec have huge variance on this dataset. However, on this clean dataset, the \EASER method leads to even higher effectiveness than the not reproducible numbers reported in the paper. High variance is also an issue for the MovieLens-1M natural noise dataset, where all simple models were more effective than DiffRec and L-DiffRec.

It can also be observed that the DiffRec variants do not offer major improvements over the simpler version of the algorithm. In particular, L-DiffRec almost always achieves similar effectiveness to DiffRec: slightly lower on the clean datasets and slightly higher on the natural noise ones. Meanwhile, T-DiffRec and LT-DiffRec are marginally better than DiffRec on the Yelp clean dataset and substantially better on the Amazon-Books clean dataset. However, as already noted, none of these variants are consistently superior to the baselines we chose. While they can be competitive on the Yelp clean dataset (albeit with high variance), many baselines are still more effective by a wide margin on Amazon-Books.

All in all, our comparison against existing baselines revealed that DiffRec (and its variants), despite its substantial computational complexity, does not lead to better effectiveness than longer-existing models. We reiterate that in our experiments, we relied on the best hyperparameters as reported in the original datasets. Clearly, we cannot rule out that the effectiveness of DiffRec can be further improved by exploring larger hyperparameter spaces, but we can confidently state that the original experimental evaluation was insufficient to support the claim that DiffRec is a competitive model.

\subsection{Collaborative Filtering Based on Diffusion Models: Unveiling the Potential of High-Order Connectivity -- CF-Diff (SIGIR '24) }
\label{subsec:CF-Diff}
\citet{DBLP:conf/sigir/HouPS24} propose CF-Diff, a method based on Gaussian diffusion with guidance. The guidance mechanism adopted by this method is based on a graph random walk and is related to the probability of reaching a user or item node after a predefined number of hops. Multi-hop information and user profiles are projected into a linear latent space and then combined through an attention mechanism.

\paragraph{Datasets}
CF-Diff is evaluated on three publicly available datasets: MovieLens-1M, Yelp, and Anime.\footnote{\url{https://www.kaggle.com/datasets/CooperUnion/anime-recommendations-database}.} Anime is a rating dataset comprising about 7.8M ratings, and is about twice as large as the Amazon-Books dataset used by~\citet{DBLP:conf/sigir/WangXFL0C23} for the evaluation of DiffRec. No particular explanation is provided regarding the selection of the dataset. No information about data preprocessing and data splitting is provided in the paper.

\paragraph{Artifacts Availability and Consistency}
A public GitHub repository\footnote{\url{https://github.com/jackfrost168/CF_Diff}} is shared by the authors. It contains the implementation of the proposed model, checkpoints, and the training-test data splits. The repository also refers to the DiffRec repository, giving credit to their implementation of the diffusion model and the evaluation code. The code for the baseline models is not present in the repository, nor is the code for optimizing and tuning the baselines. A further analysis of the provided artifacts revealed that the provided material is \emph{not complete} and in several ways \emph{not fully consistent} with the descriptions in the paper.

In terms of the provided data, we find that the MovieLens-1M dataset has been preprocessed and split as for DiffRec, with statistics matching those reported in the paper, though this is not explicitly stated. It is unclear whether the Yelp and Anime datasets were split and preprocessed using the same methodology, as the code for data splitting is not available. This is particularly relevant since the shared datasets are significantly smaller than their original versions. While the Yelp statistics reported in the paper match those reported in the DiffRec paper, the results reported are quite different, suggesting the two splits are not consistent. We recall that the data were split temporally for the DiffRec paper, but it is unclear if the same was done for the evaluation of CF-Diff. Another observation in this context is that the shared validation split includes more than 1,000 cold users. Furthermore, the statistics of the shared Anime dataset differ from both the values reported in the paper and those found on the Kaggle dataset page. Specifically, the shared version of the Anime dataset has only half the reported number of items, 25\% fewer users, and 200,000 fewer interactions.
Table~\ref{tab:CF-Diff_dataset_split_stats} shows the statistics for the shared dataset splits for the Yelp and Anime datasets and the numbers provided in the paper.

\begin{table}[h]
\resizebox{\linewidth}{!}{%
    \begin{minipage}{\textwidth}
    \centering
    \footnotesize
    \begin{tabular}{llll}
    \toprule
    Dataset			& & CF-Diff shared splits	& Reported in CF-Diff and DiffRec papers \\
    \midrule
    Yelp 2018       & Interactions  & 1,561,406 & 1,402,736 \\
                    & Users         & 31,668    & 54,574    \\
                    & Items         & 38,048    & 34,395    \\
    \midrule
    Anime           & Interactions  & 7,634,579 & 7,813,737 \\
                    & Users         & 6,969      & 11,200 \\
                    & Items         & 54,190     & 73,515 \\
	\bottomrule
   	\end{tabular}
    \end{minipage}
    }
    \caption{Datasets characteristics of shared splits and numbers provided in the paper.}
    \label{tab:CF-Diff_dataset_split_stats}
\end{table}

Another inconsistency involves the guidance data. The guidance is precomputed and shared, along with a code snippet for computing it for the case of 2-hop connections. However, this code does not support computing multi-hop guidance for more than 2 hops, even though the paper states the model optimization explored 4 hops as well. More importantly, the shared code and guidance data appear inconsistent with the explanations provided in the paper for several reasons:
    \begin{enumerate*}
        \item the normalization constant $N_{h-1,h}$ for the random walk probability is user-dependent in the paper but is the same for all users in the code;
        \item while the second-hop guidance should represent a probability distribution over the users, the code computes a probability distribution over the items;
        \item the paper contains an example of the guidance mechanism in which the nodes visited in the previous hop are disconnected from the graph by removing the edges connecting to them so that the random walk of a user will not traverse the same node twice. However, the provided implementation does not include this mechanism.
    \end{enumerate*}

Additionally, some model components are not described in the paper. For instance, the multi-hop guidance is initially fed into the same encoder as the user profile before being processed by specific multi-hop encoders. Other unmentioned components include multiple normalization layers (after encoding the user profile and attention layers), dropout layers (after each normalization and attention layer), activation functions, and some aggregations that are not explicitly described. For example, certain aggregations involve weighted sums of intermediate embeddings with fixed weights defined in the code. It is unclear whether these constants were tuned.

Lastly, the loss function reported in the paper appears to differ from the one actually minimized during training. The paper describes the complete Variational Lower Bound (VLB), but the implementation follows the same approximation as DiffRec \cite{DBLP:conf/sigir/WangXFL0C23}, where the loss is computed only for a single sampled time step $t$ rather than for all time steps.

\paragraph{Methodological Issues}
The proposed model is benchmarked against nine baselines, including five non-generative and four generative ones, with the latter group including DiffRec. While the evaluation code is based on DiffRec, as stated above, the set of baselines partially overlaps but is not the same, \eg it does not include a traditional matrix factorization model, and also the set of neural models is not identical. No explanation regarding the choice of the baseline models is provided.

The paper provides details about the explored hyperparameter ranges only for the proposed model, but not the optimal values for each dataset.\footnote{For the number of hops, a range from 1 to 4 is reported. The provided code, as mentioned, only includes a snippet for 2-hop connections.} Regarding the hyperparameters of the diffusion model, the authors mention that they ``essentially'' follow the settings of DiffRec. Besides its vague nature, this statement is surprising considering that DiffRec had not been tuned or evaluated on the Anime dataset in~\cite{DBLP:conf/sigir/WangXFL0C23}.

No information about hyperparameter tuning for the baselines is provided in the paper. This information cannot be found in sufficiently detailed form in the shared GitHub repository either. As a result, no conclusions can be drawn from the provided materials about the effectiveness of the baseline optimization process.

\begin{table}[ht]
\centering
\footnotesize
\begin{tabular}{l|cc|cc}
\toprule
\textbf{Dataset} & \multicolumn{2}{c}{\textbf{Compatible}} & \multicolumn{2}{c}{\textbf{Not Compatible}} \\
& \textbf{Stable} & \textbf{Unstable} & \textbf{Stable} & \textbf{Unstable} \\
\hline
MovieLens-1M (clean) & 0/4 & 0/4 & 0/4 & 4/4 \\
Yelp (clean)         & 1/4 & 0/4 & 3/4 & 0/4 \\
Anime                & 0/4 & 4/4 & 0/4 & 0/4 \\
\hline
\textbf{Total}       & \textbf{1/12} & \textbf{4/12} & \textbf{3/12} & \textbf{4/12}\\
\bottomrule
\end{tabular}
\caption{Reproducibility outcomes of CF-Diff.}
\label{tab:summary-cfdiff}
\end{table}

\paragraph{Reproducibility}
We were able to execute the provided code for all three datasets. However, there was only one case, the NDCG@20 metric on the Yelp dataset, where we could obtain reproducible results \idest \emph{compatible} and \emph{stable}. In all other cases, one of the reproducibility criteria (either stability or compatibility) was violated. Overall, from these observations we might conclude that only a small part of the originally reported numbers could be reproduced. As we will see later, however, there seem to be substantial issues with the Yelp dataset that will lead us to the conclusion that none of the results of CF-Diff can be reliably reproduced.

The results obtained for the Anime dataset are \emph{partially reproducible} because, while the results reported in the paper are \emph{compatible} with our experiments (NDCG@10 of $0.4677 \pm 0.0627$ with the original paper reporting $0.5152$), the results are highly \emph{unstable} with the variance reaching up to 13\% of the mean for NDCG@10 and 15\% for Recall@10. On the MovieLens-1M dataset, the results are clearly \emph{non reproducible} because they are both \emph{not compatible} \idest falling outside our tolerance interval (NDCG@10 of $0.0851 \pm 0.0030$ with the original paper reporting $0.0912$), and \emph{unstable}, \idest exhibiting relatively high variance.
On the Yelp dataset, most results reported in the original paper are \emph{not compatible}, being significantly outside the tolerance intervals, and we obtain much better results for some metrics (NDCG@10 of $0.0401 \pm 0.0004$ with the original paper reporting $0.0368$).

Table~\ref{tab:CF-Diff-yelp2018_original-result} presents the results for the Yelp dataset as an example, since we observed particularly strong deviations from the numbers reported in the paper for this dataset, with a 40\% difference in Recall compared to our estimated mean. Again, we only highlight one set of results here for the sake of conciseness. Table \ref{tab:summary-cfdiff} summarizes the reproducibility outcomes on all the experiments executed. The complete experimental results for all datasets are provided in Appendix~\ref{app:cfdiff}.

\tableArticleResult{CFDiff}{CF-Diff}{yelp2018_original}{Yelp (clean)}{1}{}{0}

\paragraph{Benchmarking Results}
Considering the results in Table~\ref{tab:CF-Diff-yelp2018_original-result}, we find that the effectiveness of CF-Diff in terms of Recall is much lower than what was reported in the original paper and that our baselines exhibit lower Recall than they did compared to the Yelp clean dataset (of similar size) shared by the authors of DiffRec. We attribute this effectiveness gap to the differences in how the dataset was preprocessed. As discussed above, it is for example unclear if the Yelp dataset was split based on a temporal ordering of the data. In contrast, we note that the NDCG values in Table~\ref{tab:CF-Diff-yelp2018_original-result} for the CF-Diff version of the Yelp dataset are often higher than for the DiffRec dataset version. This supports our hypothesis that the datasets have been preprocessed in different ways, making a comparison of absolute values across datasets impossible. As such, we conclude that the work is entirely not reproducible, even though the obtained NDCG values are in one case comparable to those reported in the paper.

Taking the actually measured effectiveness values for CF-Diff as a reference, we see from Table~\ref{tab:CF-Diff-yelp2018_original-result} that all baselines, except for \MFBPR, outperform CF-DiffRec on both metrics and on both cut-off lengths. While the difference between CF-DiffRec and traditional nearest-neighbors methods is rather small, the gap between CF-DiffRec and the best-performing models (GF-CF and MultVAE) on this dataset, is substantial.

Looking at the results for the Anime dataset, the effectiveness of CF-Diff is roughly in the range and slightly higher than that of the UserKNN method. Again, however, many of our baselines outperform CF-Diff on this dataset, with  GF-CF and MultVAE again exhibiting solid effectiveness gains. Notably, \EASER and SLIM reach even higher effectiveness on the Anime dataset.

For the MovieLens-1M dataset, we observed strong variance in effectiveness across several model executions of CF-Diff, making the results not reproducible. If we take the mean of several runs of CF-Diff, we again find the effectiveness of the model in the range of UserKNN, but usually a bit lower. UserKNN is generally quite competitive on this dataset, though. Other models like \EASER, SLIM, or MultVAE, however, outperform UserKNN as well (and thus also CF-Diff).

Overall, we found that the results reported for CF-Diff in the original paper are not reproducible according to our criteria. The model often exhibits strong variance across multiple execution runs. Several issues related to the shared datasets are further barriers to reproducibility. Finally, when comparing the measurements obtained by executing the shared artifacts, we find that CF-Diff is outperformed by many existing baselines from the literature.

\subsection{Graph Diffusion Model for Collaborative Filtering -- GiffCF (SIGIR '24) }
\label{subsec:GiffCF}
\citet{DBLP:conf/sigir/ZhuWZX24} propose GiffCF, a method based on diffusion models and graph smoothing \cite{DBLP:conf/sigir/0001DWLZ020, DBLP:conf/cikm/ShenWZSZLL21}. User interactions are interpreted as a signal on the item-item graph. The forward process employs a newly proposed graph filter to smooth the signal and corrupt user profiles. The backward process reconstructs the corrupted user profile through a two-stage neural network that takes the corrupted profile as input, along with two additional vectors as guidance: the original user profile and its smoothed version. During training, random dropout is applied to the original user profile used as guidance.

\paragraph{Datasets}
Like DiffRec, GiffCF was evaluated on the following three datasets: MovieLens-1M, Yelp, and Amazon-Books. The authors rely on the datasets, preprocessing, and data splitting procedures that were used for DiffRec paper. The corresponding train, validation and test splits of the ``clean'' datasets of DiffRec are re-shared. A 7:1:2 data splitting ratio is mentioned in the paper. We recall, however, that the ratio of the shared splits for the DiffRec model is rather close to 7:2:1.

\paragraph{Artifacts Availability and Consistency}
A public GitHub repository\footnote{\url{https://github.com/VinciZhu/GiffCF}} shared by the authors
contains the model implementation, checkpoints, and training-test data splits. However, as with DiffRec, the repository does not include the code for data splitting and preprocessing, implementations of the baseline models, or the source code used for training and optimizing them. Additionally, the paper does not provide sufficient details on the tuning procedures for most baselines. Some hyperparameters are fixed without clear justification, while others are either \emph{``tuned or set to the default values suggested in the original papers.''}
Overall, the provided artifacts are \emph{consistent} with the descriptions in the paper, except for the data splitting ratio, but \emph{not complete}, as several artifacts are missing to ensure full reproducibility.

\paragraph{Methodological Issues}
The proposed model is benchmarked against eight baselines from the literature, which also include DiffRec and L-DiffRec as previous diffusion-based models. While GiffCF builds on the results of DiffRec, the other baselines are only partially overlapping with the baselines used in the DiffRec paper. A matrix factorization (MF) model is included which is said to optimize the BPR loss, however the provided reference points to a different MF model. From the remaining five models only two, MultVAE and LightCGN, overlap with the baselines from the DiffRec paper. No particular justification is provided regarding the choice of the baselines.

Some details regarding the hyperparameters of the proposed model and the baselines are reported, but how those values were obtained is not explained. Worryingly, for some baselines the authors report that they used a fixed value of 200 across models and datasets for central hyperparameters like the user and item embedding size. However, model features like the embedding size are crucial hyperparameters that can significantly impact the effectiveness of a model~\cite{shehzad2023everyone}. Furthermore, some hyperparameter values are reported to be chosen based on the values of the original papers in which they were proposed. It is however not clear if the same datasets and evaluation protocol were used in the original papers. In sum, these choices indicate that the baselines were not systematically tuned for all datasets. The source code does not provide a sufficiently detailed description of the optimization protocol and hyperparameter search spaces are not provided either.

For the DiffRec and L-DiffRec baselines, no hyperparameter tuning was done, and they were evaluated using the checkpoints provided in the original paper. While this at first appears reasonable given that both papers use the same data splits, an inspection of the code revealed that these baselines were optimized using a different formulation of the Recall metric, with DiffRec \cite{DBLP:conf/sigir/WangXFL0C23} using the formulation where the denominator is the number of interactions in that user's ground truth, while GiffCF is optimized using the \emph{normalized} version of Recall where the denominator is the minimum between that value and the recommendation list length. This discrepancy will make the results of the two methods not comparable under Recall.

An alarming methodological issue concerns the hyperparameter tuning strategy for the proposed model. The paper states: ``\emph{For each trainable method, we use the validation set to select the best epoch and the testing set to tune the hyper-parameters and get the final results.''} This statement suggests that hyperparameters are optimized on the same set used for reporting the final results, introducing data leakage. If this is truly the case,  this represents a major flaw, as it invalidates effectiveness comparisons and artificially inflates the reported effectiveness of the method.
As mentioned earlier, a previous study on evaluation practices in recommender systems research unfortunately revealed that hyperparameter tuning on test data is actually not uncommon, and about 37\% of the papers analyzed in~\cite{Lv2021ProgressHGN} were based on such a methodologically flawed research design.

A minor issue concerns the early-stopping procedure. The source code revealed that different patience values and evaluation frequencies are used for different datasets. While this may be a valid choice, it suggests that these hyperparameters were tuned, yet this process is not described in the paper.

\begin{table}[ht]
\centering
\footnotesize
\begin{tabular}{l|cc|cc}
\toprule
\textbf{Dataset} & \multicolumn{2}{c}{\textbf{Compatible}} & \multicolumn{2}{c}{\textbf{Not Compatible}} \\
& \textbf{Stable} & \textbf{Unstable} & \textbf{Stable} & \textbf{Unstable} \\
\hline
MovieLens-1M (clean) & 0/6 & 4/6 & 0/6 & 2/6 \\
Yelp (clean)         & 0/6 & 0/6 & 0/6 & 6/6 \\
Amazon-Books (clean) & 1/6 & 0/6 & 0/6 & 5/6 \\
\hline
\textbf{Total}       & \textbf{1/18} & \textbf{4/18} & \textbf{0/18} & \textbf{13/18} \\
\bottomrule
\end{tabular}
\caption{Reproducibility outcomes of GiffCF.}
\label{tab:summary-giffcf}
\end{table}

\paragraph{Reproducibility}
We were able to run the code artifacts provided by the authors in the public GitHub repository. An inspection of the results however showed that the effectiveness of GiffCF, as reported in the original paper, could not be reproduced, except for the one single measurement (Recall@20 on the Amazon-Books dataset), see Table \ref{tab:summary-giffcf}.

Table \ref{tab:GiffCF-ml-1m_clean_original-result} presents the results for the MovieLens-1M dataset as an example, since we observed significant variability in our results estimates.
Indeed, on MovieLens-1M, the original paper's results are \emph{partially reproducible} because, while they are mostly \emph{compatible}, \idest fall within our estimated tolerance interval, they are highly \emph{unstable}, exhibiting a variance as high as 14\% of the mean on NDCG (NDCG@10 of $0.0842 \pm 0.0121$ with the original paper reporting $0.0962$) and even 18\% on Recall@10. Except for one single measurement on Amazon-Books, neither of the reproducibility criteria are satisfied for the Yelp and Amazon-Books datasets. For both datasets, all results reported in the original paper were \emph{not compatible} and \emph{unstable}, with the variance itself being rather large (2.5 - 3\% of the mean).
Table \ref{tab:summary-giffcf} summarizes the reproducibility outcomes for all executed experiments. The complete experimental results are provided in Appendix \ref{app:giffcf}.

\tableArticleResult{GiffCF}{GiffCF}{ml-1m_clean_original}{MovieLens-1M (clean)}{0}{}{0}

\paragraph{Benchmarking Results}
As shown in Table~\ref{tab:GiffCF-ml-1m_clean_original-result}, in our experiments the effectiveness values of GiffCF on the MovieLens-1M dataset are in the range of the GF-CF and the MF methods. These two methods are however both among the weakest performing baselines in our comparison. The best models for this dataset include \EASER, SLIM and \pbeta, and UserKNN  which all outperform GiffCF. Since the paper states that the test data were used for hyperparameter optimization, we also report the result of our own hyperparameter optimization of GiffCF. The results indicate that the effectiveness of GiffCF drops consistently across all measurements and exhibits increased variance, putting it even further behind the baselines. Clearly, optimizing on test data causes information leakage that artificially inflates the effectiveness of a method. Unfortunately, on the other datasets GiffCF frequently required more than the 24 GB available on our GPU and we were not able to conduct a new hyperparameter optimization, but there is no reason to doubt that the result would be a drop in effectiveness for GiffCF on both of them.

A similar picture can be observed for the Amazon-Books dataset, where the mean of the observed values for GiffCF is much lower than most of our baselines. On this dataset, the ItemKNN method is again particularly strong, and almost consistently leads to higher effectiveness than those reported, yet not reproducible results reported in the GiffRec paper.

On the Yelp dataset, finally, the effectiveness for GiffCF in our experiments is competitive with our baselines and GiffCF outperforms several of them. However, GiffCF falls behind some of the baseline models, including MultVAE and GF-CF. We note that the results reported in the original paper would have been competitive or superior to all our baselines for this dataset, but we could not reproduce these results, and the various methodological issues notably the optimization on the test data, invalidate the original measurements.

In summary, the analysis of the GiffCF paper and the shared resources revealed a number of open methodological issues. Furthermore, the model leads to high variance in the results for most datasets, and we were unable to reproduce the numbers that were reported in the paper.

\subsection{Denoising Diffusion Recommender Model -- DDRM (SIGIR '24)}
\label{subsec:DDRM}
\citet{DBLP:conf/sigir/ZhaoWXSFC24} proposed DDRM, a diffusion-based recommender model designed to denoise pretrained user and item embeddings derived from a backbone model. The backbone models considered are LightGCN, SGL \cite{DBLP:conf/sigir/WuWF0CLX21}, and matrix factorization with BPR \cite{DBLP:conf/uai/RendleFGS09}. The loss function combines the Bayesian Personalized Ranking (BPR) \cite{DBLP:conf/uai/RendleFGS09} loss and a Variational Lower Bound (VLB) \cite{DBLP:conf/nips/KingmaSPH21} term. Gaussian noise is iteratively added to the user and positive item embeddings, which are then recovered by separate denoisers for users and items. The item denoiser uses the original user embedding as guidance, while the user denoiser uses the original positive item embedding. During inference, the average embedding of historically liked items is treated as an item embedding and fed into the item diffusion process. The denoised embedding is then used to score items and generate the recommendation list.

\paragraph{Datasets}
The evaluation of DDRM, like CF-Diff, was based on the three datasets from the DiffRec paper: MovieLens-1M, Yelp, and Amazon-Books. The same data splitting and preprocessing strategies as the ``natural noise'' and ``random noise'' versions of the DiffRec paper were used. The ``clean'' version of the datasets was not used because the focus of the original evaluation is to denoise implicit feedback. A data splitting ratio of 7:1:2 is reported in the paper.

\paragraph{Artifacts Availability and Consistency}
The public GitHub repository\footnote{\url{https://github.com/Polaris-JZ/DDRM}} shared by the authors contains a working implementation of the method. However, data splits are only available for the ``natural noise'' version of the datasets. As with DiffRec, it was not possible to run experiments on the ``random noise'' dataset versions, since neither the data splits nor the code for splitting and preprocessing the data are provided.\footnote{The question of how this experiment could be included in the paper proposing DDRM, given that it is supposed to use DiffRec ``random noise'' splits which were not among the available artifacts, may be answered by noting that most authors of DDRM~\cite{DBLP:conf/sigir/ZhaoWXSFC24} are also authors of DiffRec~\cite{DBLP:conf/sigir/WangXFL0C23}.}

A complete execution of the experiments described in the paper also requires embeddings of each backbone model for each dataset. This requirement was originally not fulfilled, since only the embeddings computed for MovieLens-1M (``natural noise'' version) were shared. Fortunately, after contacting the authors by email, they updated the GitHub repository with the embeddings of all backbone models for each dataset. However, no details are provided about how the backbone models were trained, and the implementation of SGL is missing.\footnote{After publishing the first version of this manuscript on arXiv, we were contacted by the authors, with whom we had several exchanges to identify the correct configuration to reproduce the results reported in the original papers and we were also provided with the original initial embeddings learned with the backbone models. While the new configuration they provided improved the results, the overall outcome of our analysis remains unchanged. In this section, we report the results obtained with the new configuration, and we provide a comparison between this configuration and the hyperparameters we originally derived from the available artifacts in Appendix \ref{app:ddrm}.}

Regarding consistency, since the data splits are the same as in DiffRec, the same issues are present. The actual ratio is closer to 7:2:1, rather than the 7:1:2 ratio reported in the paper. Additionally, the Yelp natural noise dataset contains a small number of overlapping interactions between the training, validation, and test sets. These interactions were removed as explained in Section \ref{subsec:DiffRec}. As with DiffRec, the implementations of the baselines and the source code used to train and optimize them are not available. Furthermore, the paper does not provide sufficient details on the baseline optimization process, since it does not specify which hyperparameters were optimized or their search ranges. This lack of information makes it difficult to assess how the baselines were tuned and to identify potential methodological issues.

Inconsistencies between the model description and the actual implementation are also present. Specifically, the paper suggests that the backbone model embeddings are directly fed into the diffusion process, but the source code indicates that they are first processed with a graph convolution before entering the diffusion process.

Moreover, the implementation of SGL is missing, even though it is listed as a backbone model. The MF implementation is present but not used (at least according to the instructions for reproducing results on MovieLens-1M).
Separate source folders exist for each backbone model, containing almost identical code, but with small variations. For example, in the SGL-based implementation, normalization layers are added, and certain mean aggregation functions are replaced by summations, while in the MF-based implementation, the loss function includes a non-linear function and an explicit regularization term that are absent from the other models (and not described in the paper).

In conclusion, the provided material is \emph{not complete} and \emph{not fully consistent} with the descriptions in the paper.

\paragraph{Methodological Issues}
DDRM was benchmarked both against four alternative state-of-the-art model-agnostic denoising models and against four other baselines. These other baselines include the DiffRec model, two other models that also served as baselines in the DiffRec paper, and an additional collaborative filtering denoising model.

Hyperparameter ranges for the proposed model are reported, but the paper does not mention how these hyperparameters were tuned and what the optimal values were. Regarding baseline optimization, it is not possible to draw conclusions on its correctness, as neither the source code nor a detailed description of the optimization protocol and hyperparameter search space is provided. The embedding size was fixed a priori across all methods to 64 ``for fair comparison.'' Again, this represents a major methodological issue, because the embedding size has to be tuned for each dataset to ensure that each baseline can reach the best possible effectiveness. By restricting the embedding size, the effectiveness of the baselines is artificially limited. Furthermore, for the hyperparameters of some backbone models, default values are chosen. However, to properly gauge the effects of denoising, all models should first be tuned to maximise their effectiveness.

A minor issue was found in the training and validation procedure. From the code, and this is not explicitly mentioned in the paper, it appears that when evaluating the model on the validation set for early stopping, only 50\% of randomly sampled users are considered. This could lead to a noisy estimate of model effectiveness on unseen data and potentially result in a suboptimal hyperparameter configuration, particularly for the number of training epochs. Nonetheless, the final evaluation on the test set is correctly performed on all users. In contrast, in our experiments, we evaluate the model on the validation set using all users, ensuring a more stable and reliable selection of the number of epochs.

\begin{table}[ht]
\centering
\footnotesize
\begin{tabular}{ll|cc|cc}
\toprule
& \textbf{Dataset} & \multicolumn{2}{c}{\textbf{Compatible}} & \multicolumn{2}{c}{\textbf{Not Compatible}} \\
& & \textbf{Stable} & \textbf{Unstable} & \textbf{Stable} & \textbf{Unstable} \\
\hline
\multirow[t]{3}{*}{DDRM-LightGCN}
  & MovieLens-1M (natural noise) & 0/4 & 0/4 & 1/4 & 3/4 \\
  & Yelp (natural noise)         & 1/4 & 0/4 & 3/4 & 0/4 \\
  & Amazon-Books (natural noise) & 0/4 & 0/4 & 0/4 & 4/4 \\
\hline
\multirow[t]{3}{*}{DDRM-MF}
  & MovieLens-1M (natural noise) & 0/4 & 4/4 & 0/4 & 0/4 \\
  & Yelp (natural noise)         & 0/4 & 0/4 & 4/4 & 0/4 \\
  & Amazon-Books (natural noise) & 0/4 & 0/4 & 4/4 & 0/4 \\
\hline
\multirow[t]{3}{*}{DDRM-SGL}
  & MovieLens-1M (natural noise) & 0/4 & 0/4 & 4/4 & 0/4 \\
  & Yelp (natural noise)         & 3/4 & 0/4 & 1/4 & 0/4 \\
  & Amazon-Books (natural noise) & 0/4 & 0/4 & 4/4 & 0/4 \\
\hline
\textbf{Total}           & & \textbf{4/36} & \textbf{4/36} & \textbf{21/36} & \textbf{7/36}  \\
\bottomrule
\end{tabular}
\caption{Reproducibility outcomes of DDRM.}
\label{tab:summary-ddrm}
\end{table}

\paragraph{Reproducibility}
The authors of DDRM share both the code and, upon request, pretrained embeddings for each backbone model (LightGCN, SGL, and MF) on the ``natural noise'' versions of the MovieLens-1M, Yelp, and Amazon-Books datasets. In order to conduct a conservative evaluation we use those backbone embeddings in our experiments. Following the original experimental protocol, we could execute the training and evaluation pipeline for each combination of dataset and backbone. Our measurements, however, could reproduce only fewer than 10\% of the results reported in the paper. Only two experiments were \emph{partially reproducible}. The first one is DDRM-SGL on the Yelp natural noise dataset, where three out of four metric values were reproduced. The second is DDRM-MF on MovieLens-1M natural noise where results were \emph{compatible}, but also highly \emph{unstable}, with a very large variance due to learning instability that leads the model to fail.\footnote{Across the 10 runs of DDRM\textendash MF on MovieLens-1M with identical hyperparameters and number of epochs, we intermittently observed a training pathology where the loss collapses to exactly zero after a few updates. When this occurs, the model yields degenerate predictions, which leads to totally wrong recommendations. These failures pull down the mean of the measured metrics and markedly increase their variance, which explains the instability we report. The collapse appears stochastic. Based on this observation and on our exchange with the original authors, we suspect that DDRM is very sensitive to numerical precision and that differences in hardware configuration may significantly affect the results.} All other experimental configurations yielded results \emph{not compatible} \idest outside of the tolerance interval, but mostly \emph{stable} indicating the existence of a systematic factor affecting the results.

Table \ref{tab:DDRM-amazon-book_noisy_original-result} reports the results on Amazon-Books (natural noise) dataset as an example, since it is the dataset with the worst results in terms of reproducibility.
With few exceptions, the variance exhibited by DDRM is generally quite low. Indeed, almost 70\% of the metrics are \emph{stable} based on our measurements, as can be noted from Table \ref{tab:summary-ddrm}. This often leads to tight tolerance intervals that make the results we obtain statistically \emph{not compatible} with those reported in the paper, despite the absolute values being relatively close to those reported in the paper \idest our estimate of DDRM-MF NDCG@10 on Amazon-Books is $0.0141 \pm 0.003$, which is very close to the value of 0.0148 reported in the paper. In this scenario, it is likely that systematic factors are playing a role in shifting the results, for example due to differences in the experimental configuration that were not explained. The complete experimental results are provided in Appendix \ref{app:ddrm}.

\tableArticleResult{DDRM}{DDRM}{amazon-book_noisy_original}{Amazon-Books (natural noise)}{1}{}{0}

\paragraph{Benchmarking Results}
Across all three datasets, the best DDRM variants do not consistently outperform the top-performing baselines. On Yelp, for instance, DDRM-LightGCN and DDRM-SGL are competitive with several neighbor-based methods but still lag behind \palpha, MultVAE and \iALS. On Amazon-Books, simpler approaches such as ItemKNN and SLIM achieve higher Recall and NDCG by a notable margin, both in our experiments and in the original ones. Similarly, on MovieLens-1M, DDRM occasionally rivals weaker baselines but remains below top methods like \EASER and SLIM.

The DDRM-MF variant is consistently the weakest among the three. While DDRM offers a sophisticated approach to collaborative filtering, its added complexity does not yield higher effectiveness compared to well-tuned classical techniques. Moreover, our findings indicate that the paper's original experiments relied on weak baselines, undermining the strength of its claims.

\section{Discussion}
\label{sec:discussion}
Having reported the outcomes of our reproducibility and benchmarking experiments in the previous section, we now critically reflect on the adoption of concepts from diffusion models in recommendation scenarios. Furthermore, we will report on the immense computational costs of using such models for recommendation tasks, and finally discuss the implications of our findings and possible future steps. We hope this discussion can inform the design of new architectures or training protocols that address the issues we identified; however, developing such methods is beyond the scope of this study.

\subsection{A Critical Assessment of DDPM for Recommendation and Future Directions}
\label{subsec:DDMP-critical}
Successfully applying an architecture to a task requires going beyond the mere intuition of its promise. When this is overlooked there is a real risk that the proposed methods will not work as intended at all and that fundamental flaws will remain undetected.\footnote{See for example a discussion on the flawed idea of applying Convolutional Neural Networks on the outer product of embeddings learned by Matrix Factorization models \cite{ferraridacremaetal2020cikm} which did not have the properties CNNs are designed to leverage.} Our finding indicate that \emph{all the papers we analyzed applied DDPMs in a way that violates some of their essential requirements and assumptions} and that \emph{none of the methods is generative in the traditional sense}, since they do not perform sampling at all. This is not a problem per se, as the field of machine learning is full of examples of successful heuristics that worked well in practice even though they are not fully theoretically understood. The problem arises when an architecture is modified in ways that break some of its fundamental assumptions, without addressing the consequences of those changes. In our study, this is compounded by the empirical observation that the resulting methods not only incur a significant computational cost, as discussed in Section~\ref{subsec:computational-cost}, but are also validated with a weak and flawed experimental protocol, undermining their claims of superior effectiveness. While our experimental results focus on collaborative filtering, and the datasets are limited to those used in the analyzed papers, we believe the fundamental issues discussed in this section are common to several applications of DDPMs in recommendation systems and extend beyond the specific methods we analyzed. We can group the issues in the following three points.

\paragraph{Keeping the Forward Process at Inference:}
Normally, as explained in Section \ref{subsec:sampling}, the forward process is used only during training to generate noisy data samples for training the denoiser model. At inference time, generation typically starts by sampling from a Gaussian distribution and then iteratively applying the denoiser model. In this sense, besides a random noise sample, there is \emph{no input data}. Clearly this means that this type of architecture \emph{is not suitable for personalized recommendation} because there is no way to constrain the generation process toward a specific user profile. This important consideration does not appear in any of the papers we analyzed, including DiffRec, which uses this very architecture \cite{DBLP:conf/sigir/WangXFL0C23}. However, even if DDPMs are not suitable for personalized recommendation, if they are able to generate plausible new user profiles they may have other applications, \eg synthetic data generation.

The papers we analyze overcome this by applying the forward process during inference as well, adding noise to the input data before applying the reverse process, similarly to what is done during training. This approach resembles traditional non-generative denoiser models, such as denoising autoencoders, where noise is injected into the user profile and a neural network learns to recover the original input. The main difference here would be that both noise injection and denoising occur over multiple steps.

\paragraph{Limited Corruption of the Input Data:}
Another indication that diffusion models are not being used as intended is the small number of diffusion steps and the limited noise scale employed in many of the analyzed papers.
From a theoretical standpoint, in DDPMs the forward process must progressively corrupt the user-item interaction data or embedding into a simple distribution (\idest a Gaussian) which becomes pure noise and should retain no information about the input data, as explained in Section \ref{par:convergence}. The reverse process, in turn, must be capable of accurately recovering this structure, possibly under the guidance of auxiliary information.
As highlighted in Section \ref{subsec:structure}, this theoretically requires very long Markov chains as forward and backward processes, so that each step can account for a very small noise perturbation, and simplify the training, as reported by \cite{DBLP:conf/icml/Sohl-DicksteinW15}: \emph{``Learning in this framework involves estimating small perturbations to a diffusion process. Estimating small perturbations is more tractable than explicitly describing the full distribution with a single, non-analytically-normalizable, potential function. [...] The longer the trajectory the smaller the diffusion rate $\beta$ can be made.''}

The number of steps used by the papers we analyze is reported in Table \ref{tab:diff-steps}. We can see how the number of steps is frequently equal to or below 10, with very few applying 50 or more. In the four papers analyzed, this limited noise injection is usually justified by the observation that if the noise scale is too large, or the Markov chain too long, the user profile input becomes too corrupted and loses its personalized information, effectively turning into Gaussian noise. However, this complete destruction of the input is a fundamental part of how diffusion models operate as it enables the sampling process and underpins their theoretical guarantees. If the model only performs well when the input is not fully destroyed, this suggests that its generative capacity is not effective for that specific task. If the denoiser operates on a partially corrupted input, the setup becomes more akin to a denoising autoencoder than a proper generative model.

\begin{table}[h]
\resizebox{\linewidth}{!}{%
    \centering
    \footnotesize
    \begin{tabular}{l|ccccccc}
	\toprule
	Algorithm	& \begin{tabular}{c}MovieLens-1M\\(clean)\end{tabular}	& \begin{tabular}{c}Yelp\\(clean)\end{tabular}		& \begin{tabular}{c}Amazon-Books\\(clean)\end{tabular}	& \begin{tabular}{c}MovieLens-1M\\(natural noise)\end{tabular}	& \begin{tabular}{c}Yelp\\(natural noise)\end{tabular}	& \begin{tabular}{c}Amazon-Books\\(natural noise)\end{tabular} 	& Anime 	\\
	\midrule
    DiffRec		& 40			& 5			& 5			& 5				& 5			& 10				& -		\\
	L-DiffRec	& 40			& 5			& 5			& 100				& 5			& 5				& -		\\
	T-DiffRec	& -			& 5			& 10			& -				& -			& -				& -		\\
	LT-DiffRec	& -			& 5			& 5			& -				& -			& -				& - 		\\
	GiffCF		& 3			& 3			& 3			& -				& -			& -				& -		\\
	CF-Diff		& 5         		& 20        		& -			& -				& -			& -				& 10        	\\
	DDRM-LightGCN	& -			& -			& -			& 50				& 30			& 50				& -		\\
	DDRM-MF		& -			& -			& -			& 20 				& 50			& 3				& -		\\
	DDRM-SGL	& -			& -			& -			& 30				& 50			& 3				& -		\\

	\bottomrule	
	\end{tabular}
    }
    \caption{Number of diffusion steps employed in each experiment.}
    \label{tab:diff-steps}
\end{table}

\paragraph{Guidance Similar to the User Profile}
The guidance signal in diffusion models is intended to constrain the sampling process to produce samples that exhibit the desired latent characteristics. As we have previously discussed, this guidance typically captures high-level, abstract features rather than specifying every low-level detail of the output. Typically there are \emph{many generated samples} that would share the same guidance signal. For example, in image generation, guidance often describes the general content or style of an image, rather than detailing the exact shape, color, and position of every object, which would restrict the space so much that only a few samples could satisfy it and it is unlikely the model would be able to learn such a distribution well.
Consider, for example, how many plausible images may correspond to a simple prompt like: \say{\emph{Draw a countryside landscape with fields, trees and farmers working with their equipment}}. There are almost infinitely many images, some of them only slightly different, that would be perfect fits for this prompt. Naturally, as the prompt becomes more specific the number of acceptable images would reduce, but the fundamental fact that a large number of them exists remains, indeed the DDPM will learn a \emph{distribution} over those images, as explained in Section \ref{subsec:structure} and Section \ref{subsec:sampling}.

Based on our previous observations on the corruption of the input data, it is clear that the only way to use DDPM in a generative way for personalized recommendation is to rely on a guidance signal $y$. However, this brings us to another mismatch between DDPM and a traditional recommendation scenario, because if we aim to use a DDPM for recommendation and we use the traditional offline evaluation, our aim is not to generate a ``good'' user profile that matches the semantics expressed in the guidance signal. Rather, we aim to generate \emph{the specific user profile} that appears in the test data (hence, that could be observed), which means that the distribution of the generation process is essentially collapsed to a single deterministic point. This seems fundamentally at odds with the architectural characteristics of DDPM.
The papers we analyze tend to use a guidance that is \emph{very specific}, such as a noisy user profile or a one-hop random walk started from the target user. In this scenario, even if the diffusion process were correctly applied at inference, starting from noise and iteratively using the denoiser, the model would still be conditioned on a partially corrupted version of the input. As a result, the entire architecture effectively becomes equivalent to a denoising autoencoder. As already observed by \citet{DBLP:conf/nips/YangWWWY023}, in the context of recommendation tasks, a \say{\emph{diffusion model is mostly used for adding noise in the training samples for robustness, and the learning objectives are largely categorized as classification instead of generation}}.
This is similar to what was observed for Generative Adversarial Networks when the \emph{condition vector} is too similar to the target user profile \cite{DBLP:conf/ecir/MaureraDC22}. To design a meaningful generative DDPM with guidance, the guidance should represent latent, high-level characteristics of the user, such that ``denoising'' from it is no longer a shortcut to retrieve a plausible output. It is only then that the model will be required to actually learn how to generate new samples, rather than merely reconstruct the input from its noisy version.

\subsubsection{Fundamental Constraints of DDPMs for Recommendation}
All of these observations suggest that, under the traditional top-n recommendation scenario, the generative ability of DDPMs offers no clear benefit and must be heavily constrained to produce reasonable results that, nevertheless, remain not competitive when compared to simple baselines under offline evaluation. We now turn to a discussion of possible reasons for this outcome, which we believe raise broader research questions relevant to the use of generative models in recommendation systems:

\paragraph{Lacking ground truth}  Diffusion models have been successfully applied in several domains, such as image and video generation, but the characteristics of these domains differ significantly from those of recommendation systems. For example, unlike image generation tasks where the training data can be considered complete and well-defined, recommendation systems are trained on inherently incomplete data. We never have access to the true underlying user preferences, only to a noisy, biased, and incomplete proxy in the form of user interaction histories. This limitation of applying diffusion models to recommendation is not addressed in any of the papers we analyze. One possible direction to address this issue could be to model the user profile as a partially corrupted version of the user's true preferences, analogous to assuming that the ground truth has already undergone a few steps of the forward process. Crucially, while some justification for using DDPMs in recommendation is based on the idea that user profiles are noisy \cite{DBLP:conf/sigir/WangXFL0C23}, this is again a case of incomplete intuitive justification that hides considerable complexity. Indeed, not all ``noise'' is the same. In many cases, what appears as noise may instead reflect behavioral patterns or systematic effects that influence how users interact with the system. Treating incomplete user profiles as merely ``noisy'', as if they had been corrupted by uniformly random noise, will conflate fundamentally different phenomena and likely result in inadequate modeling. Therefore, implementing this idea in practice would require to account for these effects. How to achieve this without introducing additional unwanted biases remains an open question.

\paragraph{Limited information structure} Compared to other fields such as image generation, where DDPMs can be trained on large corpora of images to learn complex patterns and structures that can be widely reused, recommendation systems provide far less rich data. In most recommendation scenarios, the available data consist of user interactions with items, typically represented by unique identifiers and a limited set of features. This provides very little information on which to base a deep understanding of the nuanced and interdependent factors that influence user behavior, such as interface design, business rules, or platform dynamics, most of which are unobservable or entirely unknown. This limitation is also reflected in the relatively low accuracy of recommendation systems, as consistently shown across research papers. One possibility is that more robust sample generation could emerge if DDPMs were trained on richer multimodal data (text, images, audio, video) allowing the model to extract more meaningful signals beyond raw identifiers. Naturally there is no guarantee this would be effective, since simply providing \emph{more data} does not necessarily mean the model will extract \emph{more information} from it. Besides, the computational cost alone would be extraordinary. Still, this points toward the potential value of developing large pretrained models for recommendation tasks.

\paragraph{Mismatching Evaluation} Generative models learn a distribution and then generate samples from it. However, in traditional offline evaluation settings, we typically only have access to a single user profile which is, effectively, one individual sample drawn from the unknown distribution that represents a combination of the user interests and many other factors (context, user interface, ...). When evaluating a generative model under this offline evaluation setting we are not assessing whether the distribution learned by the DDPM is \emph{reasonable} or capable of generating diverse, high-quality user profiles. Instead, we evaluate whether the model can recreate \emph{the specific user profile} used for training or inference. This may be the key reason behind the inconsistencies between the formal definition of DDPMs and their practical use in the papers we analyze, that is, constraining as much as possible the generative capacity of DDPMs to stay close to the original user profile. Devising evaluation strategies that reflect the real-world effectiveness of generative models while also accounting for their unique characteristics remains a challenge across many fields \cite{DBLP:conf/icml/AlaaBSS22,DBLP:conf/nips/SteinCHSRVLCTL23}. In recommendation, this challenge is compounded by long-standing issues with offline evaluation, which is known to be a weak proxy for actual user satisfaction especially when the differences in effectiveness between models are small. There is a pressing need to develop and validate new evaluation methodologies, ideally supported by online experiments. The community's continued reliance on a narrow set of accuracy metrics such as Recall and NDCG is already recognized as insufficient, providing only an incomplete view of the model's behavior in practice. In recent years, there has been growing awareness of the importance of evaluating additional dimensions such as diversity, fairness, and novelty. As generative models become more complex, it is likely that new dimensions will emerge and that an evaluation pipeline focused solely on accuracy metrics will become inadequate to assess how such models will actually perform in practice.

Overall, none of the papers we analyzed were able to demonstrate that DDPMs are strong candidates for recommendation systems. This appears to be the result of several contributing factors, some of which may be fundamental and could require substantial research effort to address. Whether there is a path to overcome those issues and limitations in the future is yet unclear.

\subsection{Computational Cost}
\label{subsec:computational-cost}

Computational efficiency is critical when deploying algorithms for large-scale or frequently updated recommender systems. Models that demand substantial resources may be infeasible for practical deployment or iterative experimentation. Therefore, for each diffusion-based model analyzed in this work, we select the largest dataset on which we ran experiments and compare its training and inference times (averaged across 10 runs) on that dataset with those of a fast, well-established baseline \idest ItemKNN, as well as with those of the most effective baseline method. If ItemKNN was already the best baseline, we used the second-best one. The experiments were conducted on a GeForce RTX 3090 GPU with 24 GB of RAM, and on an i9-13900K CPU with 64 GB of RAM (see Section \ref{subsec:computational_resources} for further details). Training time is the time needed to train the model with the best hyperparameters (averaged across 10 runs for diffusion-based models), inference time refers to the total time required to serve recommendations to all users in the test set, and throughput is calculated as the number of test users divided by the total inference time.

Table \ref{tab:time} presents the results. Overall, diffusion-based methods tend to have longer or more variable training times than ItemKNN, but they often provide faster inference, on par with or better than linear methods such as SLIM. This fast inference time can be explained by the relatively small number of diffusion steps that most methods adopt, see Table \ref{tab:diff-steps}. However, the exact trade-off between time-cost and model effectiveness varies depending on the dataset. For instance, T-DiffRec and LT-DiffRec require nearly 10,000 sec of GPU time for training on Amazon-Books, whereas CF-Diff is more moderate (about 550 sec) on Anime but has a less efficient inference than SLIM.

In some cases, particularly on \emph{Amazon-Books clean}, ItemKNN excels in both effectiveness and efficiency. Ultimately, our analysis highlights the need to evaluate the full cost–benefit ratio, since even the most effective diffusion-based methods may introduce heavy training overheads, and simpler algorithms frequently provide a more favorable trade-off between effectiveness and efficiency.

\begin{table}[th!]
    \caption{Comparison of training and inference time of the diffusion-based models, ItemKNN, and the most effective baseline method (if ItemKNN is already the most effective baseline, we chose the second best one). For each dataset, the most effective algorithm, as well as its timings, are underlined. The best values for training time, inference time and throughput are highlighted in bold.}
    \label{tab:time}
    \footnotesize
    \centering
    \begin{tabular}{llllc|}
    \toprule
    Dataset & Algorithm & \begin{tabular}{@{}c@{}}Train Time\\$[\mathrm{sec}]\ (\downarrow)$\end{tabular} & \begin{tabular}{@{}c@{}}Inference Time\\$[\mathrm{sec}]\ (\downarrow)$\end{tabular} & \begin{tabular}{@{}c@{}}Throughput\\$[\mathrm{users}/\mathrm{sec}]\ (\uparrow)$\end{tabular} \\
    \midrule
    Amazon-Books (clean)  & DiffRec       & 2002.87 $\pm$ 1678.00    & 281.74 $\pm$ 18.26   & 386   \\
                        & L-DiffRec     & 9233.08 $\pm$ 5298.20    & \textbf{266.23 $\pm$ 10.35}   & \textbf{409}   \\
                        & T-DiffRec     & 9072.48 $\pm$ 3896.28    & 311.76 $\pm$ 24.33   & 349   \\
                        & LT-DiffRec    & 9707.24 $\pm$ 7387.88    & 266.52 $\pm$ 12.35   & 408   \\
                        & \underline{ItemKNN}      & \underline{\textbf{71.35}}                    & \underline{627.40}               & \underline{173}   \\
                        & SLIM          & 40181.91                 & 312.86               & 348   \\
    \midrule
    Anime               & CF-Diff   & 553.09 $\pm$ 514.80  & 233.06 $\pm$ 25.45   & 233   \\
                        & ItemKNN  & \textbf{11.65}                & 321.06               & 169   \\
                        & \underline{SLIM}      & \underline{3186.15}              & \underline{\textbf{225.26}}               & \underline{\textbf{241}}   \\
    \midrule
    Amazon-Books (clean)  & GiffCF    & 10412.45 $\pm$ 5005.03   & 548.19 $\pm$ 19.22   & 199   \\
                        & \underline{ItemKNN}  & \underline{\textbf{23.36}}                    & \underline{\textbf{128.66}}               & \underline{\textbf{846}}   \\
                        & SLIM      & 9308.63                  & 153.85               & 707   \\
    \midrule
    Amazon-Books (natural noise) & DDRM-LightGCN & 459.34 $\pm$ 1192.94     & 337.38 $\pm$ 4.88    & 323   \\
                        & DDRM-MF       & 682.57 $\pm$ 1240.67     & 319.52 $\pm$ 4.42    & 341   \\
                        & DDRM-SGL      & 1557.23 $\pm$ 1430.76    & 327.63 $\pm$ 3.08    & 332   \\
                        & \underline{ItemKNN}      & \underline{183.05}                   & \underline{571.37}               & \underline{190}   \\
                        & SLIM          & 44146.82                 & 529.56               & 205   \\

    \bottomrule
    \end{tabular}
\end{table}

\subsection{Implications and Outlook}
\label{sec:Discussion}
Overall, the results of our analyses are sobering. Ultimately, the progress that is actually achieved by the proposed diffusion recommender models compared to existing ones on the collaborative filtering task we focused on remains unclear, despite the substantial computational cost and carbon footprint of these models.

According to our analyses, two main factors that are apparently preventing us as a research community to make substantial progress seem to persist, without signs of a major change on the horizon~\cite{cremonesi2021aimag}:
\begin{enumerate}
    \item \emph{Reproducibility Crisis:} While researchers increasingly share the code of their newly proposed models, reproducibility remains an issue. On the one hand, the shared artifacts are often incomplete and in particular the code for the tuning and running compared baseline models is almost always missing. On the other hand, the provided models and scripts often either do not lead to the results reported in the paper, or they exhibit a large variance across several execution runs.
    \item \emph{Methodology Crisis:} It is easy to agree that nothing can be concluded from a comparative experiment in machine learning in which one model is meticulously tuned whereas other models are barely tuned or not properly tuned at all~\cite{shehzad2023everyone}. Still, such an approach seems to be widely accepted in the research community.\footnote{We reiterate that we do not claim that the baselines in the studied papers were not properly tuned. We can only report that little to no information about this process is documented in the papers or available in the provided code repositories and that our simple but properly tuned baselines are much more effective.}
\end{enumerate}

Overall, we find that several of the methodological issues and common practices in recommender systems research that we reported in depth in Section~\ref{sec:previous-studies} are also found in latest research publications, hampering both reproducibility and progress. Specifically, we also found that it is still uncommon for researchers to share artifacts that are suited to reproduce the \emph{entire experiment} that was conducted to support the common claim of advancing the state-of-the-art. Typically, ensuring complete reproducibility would require providing all necessary materials include the code for the proposed model \emph{and} the baselines, hyperparameter ranges, hyperparameter optimization strategies, or data preprocessing code~\cite{Shezad2025WeShare}. In addition to these common limitations, our present work also points to an underlying conceptual mismatch when researchers adopt a method that was designed for a quite different task to recommendation problems.

Looking forward, we observe that reproducibility problems are increasingly acknowledged and addressed in the research community. Sharing code is nowadays much more common than it was ten or fifteen years ago. Furthermore, various conferences and journals nowadays include reproducibility as an evaluation criterion in their review process. Several conferences also run dedicated reproducibility tracks, which are fully devoted to these pressing issues. In addition, various reproducibility checklists have been put forward, both in the area of recommender systems~\cite{Lops2023}, in machine learning~\cite{JMLR:v22:20-303}, and in AI in general.\footnote{See, for example, the guidelines of the AAAI conference: \url{https://aaai.org/conference/aaai/aaai-23/reproducibility-checklist/}} Clearly, reproducibility issues as reported in our paper are not unique to applied areas like recommender systems, but can also be identified in the broader fields of machine learning and AI research~\cite{Semmelrock2024Repro,Gunderson2022sources,Gunderson2018State}.

Besides the more thorough consideration of such reproducibility guidelines when evaluating scientific papers before publication, we believe that it is equally important to further raise awareness in the community about the sustained problems it has. If we look at the work by Konstan and Adomavicius~\cite{Konstan2013Toward} mentioned earlier, a number of methodological issues of today were already identified many years ago in 2013. In the summary table of methodological issues in~\cite{Konstan2013Toward}, the authors, for example, mention poorly described evaluation setups, comparisons with weak baselines, or missing details regarding the configuration and tuning of the baselines. \citet{bauer2023overcoming} more recently therefore argue that further \emph{educating} the various actors in the academic research and publication ecosystem of these long-standing issues will be a key factor to avoid similar problems in the future. These actors not only include researchers and reviewers, but also educators, scholars who for example act as journal editors or evaluators of grant proposals, as well as industry practitioners.

From a technical perspective, ensuring exact reproducibility should in principle be easy to establish for many forms of algorithms research in recommender systems. Most published research is based at least partially on publicly available datasets and on freely available machine learning libraries. As such, there are no strong technological barriers. With the help of cloud-based services, virtualization or containerization technology, it is nowadays relatively easy to share entire execution environments with other researchers which enable ``one-click'' reproducibility. However, it is very rare that researchers share environments where experimental code can be easily executed and outputs the numbers that are reported in the papers both for the newly proposed models and the baseline models.

We believe that this phenomenon can be attributed to today's research and publication culture in the field of recommender systems and applied machine learning in general. This culture is largely focused on conferences, competitive, fast-paced, and deadline-driven, as noted, for example, by Turing laureate Yoshua Bengio in~\cite{bengio2020timetorethink}. As a result, there are apparently not enough incentives for scholars to invest the substantial amount of effort that is needed to prepare easy-to-use reproducibility packages and in-depth documentation material. On the contrary, for many publication venues the provision of any form of reproducibility artifacts is still optional. The provision of such artifacts is maybe increasingly considered as a positive aspect by reviewers, but as our study shows, the provided materials are in many cases incomplete and do not allow for full reproducibility.

\section{Concluding Remarks}
\label{sec:concluding-remarks}
In their seminal work on the evaluation of recommender systems from 2004, \citet{Herlocker2004Evaluating} speculated \emph{``that algorithms trying to make better predictions on movie datasets may have reached the optimal level of error given human variability.''}. Today's research certainly does not focus on rating prediction anymore and is also  not limited to the movie domain. Some results presented in this paper and in earlier reproducibility works like~\cite{ludewiglatifiumuai2020} however indicate that the last twenty years of research may not have brought us as much progress as we would expect from the published literature. In our present work as well as in many previous works as summarized in Section \ref{sec:previous-studies}, we find indications that the research community still seems to struggle to adopt rigorous methodological approaches to ensure reliable progress in terms of algorithms research.

\color{black}

\section{Acknowledgements}
We acknowledge ISCRA for awarding this project access to the LEONARDO supercomputer, owned by the EuroHPC Joint Undertaking, hosted by CINECA (Italy).

\appendix

\bibliographystyle{ACM-Reference-Format}
\bibliography{references}

\clearpage

\appendix
\section{Complete Experimental Results}
\label{app:complete-results}

\subsection{Full List of Baselines}
The full set of baselines we used in our experiments includes 15 collaborative algorithms, most of which were also included in earlier reproducibility studies as well~\cite{ferraridacrema2020tois,ferraridacremaetal2020cikm}:

\paragraph{Non-personalized Approaches}
\begin{itemize}
    \item \textbf{Random}: A non-personalized method recommending random items the user has not yet interacted with.
    \item \textbf{TopPop}: A non-personalized method recommending the most popular items to all users that the user has not yet interacted with.
    \item \textbf{Global Effects}: A non-personalized method where the item score is the average of all ratings it has received. The denominator includes a constant term to penalize the scores of items with low support.
\end{itemize}

\paragraph{Collaborative Nearest-Neighbor Techniques}
\begin{itemize}
    \item \textbf{UserKNN}: A user-based nearest-neighbor algorithm~\cite{DBLP:conf/cscw/ResnickISBR94}, with cosine similarity and shrinkage~\cite{bell2007improved}.
    \item \textbf{ItemKNN}: An item-based nearest-neighbor algorithm~\cite{DBLP:conf/www/SarwarKKR01}, with cosine similarity and shrinkage~\cite{bell2007improved}.
\end{itemize}

\paragraph{Graph-based Approaches}
\begin{itemize}
    \item \textbf{\palpha}: A graph-based method modeling a random walk on the bipartite graph of user-item interactions.
    \item \textbf{\pbeta}: A graph-based method that uses a two-step random walk from users to items and vice versa, where transition probabilities are computed from the normalized ratings~\cite{DBLP:journals/tiis/PaudelCNB17}.
    \item \textbf{GF-CF}: A graph-based method that is based on a low-pass filter and has a closed form solution~\cite{DBLP:conf/cikm/ShenWZSZLL21}.
    \item \textbf{LightGCN}: A graph-based method that performs message passing on the user-item adjacency matrix to learn user and item embeddings with BPR~\cite{DBLP:conf/sigir/0001DWLZ020}.
\end{itemize}

\paragraph{Item-Based Machine Learning}
\begin{itemize}
    \item \textbf{\EASER}: An ``embarrassingly shallow'' linear model with strong connections with autoencoders and a closed form solution~\cite{DBLP:conf/www/Steck19}.
    \item \textbf{SLIM}: An item-based model that uses linear regression and the ElasticNet loss to compute the item similarity~\cite{DBLP:conf/icdm/NingK11}.
    \item \textbf{SLIM-BPR}: An item-based model similar to SLIM that computes the item similarity optimizing the \emph{Bayesian Personalized Ranking} (BPR) loss~\cite{DBLP:conf/uai/RendleFGS09}.
\end{itemize}

\paragraph{Matrix Factorization Techniques}
\begin{itemize}
    \item \textbf{\MFBPR}: A matrix factorization method based on the \emph{Bayesian Personalized Ranking} (BPR) loss~\cite{DBLP:conf/uai/RendleFGS09}.
    \item \textbf{MF-WARP}: A matrix factorization method based on the \emph{Weighted Approximate-Rank Pairwise} loss (WARP).
    \item \textbf{SVDpp}: A matrix factorization method for rating prediction accounting for user biases~\cite{DBLP:conf/recsys/LercheJ14}.\footnote{Note that to adapt SVDpp for the task of top-n recommendation we sample a certain quota of missing interactions and attribute them a rating of zero. The specific quota is a hyperparameter.}
    \item \textbf{PureSVD}: A matrix factorization method based on the truncated SVD decomposition of the user-item interaction matrix~\cite{DBLP:conf/recsys/CremonesiKT10}.\footnote{We use a standard SVD decomposition method provided in the \small{\texttt{scikit-learn}}  \footnotesize package for Python.}
    \item \textbf{\iALS}: A matrix factorization method for ranking tasks based on alternating least-squares~\cite{DBLP:conf/icdm/HuKV08}.

\end{itemize}

\paragraph{Autoencoders}
\begin{itemize}
    \item \textbf{MultVAE}: A variational autoencoder that assumes a multinomial likelihood for user-item interactions \cite{DBLP:conf/www/LiangKHJ18}.
\end{itemize}

The hyperparameter ranges and distributions are the same used in \cite{ferraridacrema2020tois,ferraridacremaetal2020cikm}, and are reported in Appendix \ref{app:baseline-hparams} for completeness.
Note that occasionally the results for \textbf{\EASER} and \textbf{SLIM-BPR} may be missing due to their memory requirements exceeding the 64 GB available on our server.

The hyperparameter values

\subsection{DiffRec}
\label{app:diffrec}

The hyperparameters for experiments labeled ``(original hparams)'' are reported in Tables \ref{tab:DiffRec-hyperparams}, \ref{tab:L-DiffRec-hyperparams}, \ref{tab:T-DiffRec-hyperparams}, and \ref{tab:LT-DiffRec-hyperparams}, recovered from the available artifacts. The hyperparameters for experiments labeled ``(new hparams)'' are reported in Table \ref{tab:new-hparams-diffrec}, as provided by the authors. Results for all datasets and baseline algorithms are reported in Table
\ref{tab:DiffRec-ml-1m_clean_original-result-appendix} for MovieLens-1M (clean),
\ref{tab:DiffRec-yelp_clean_original-result-appendix} for Yelp (clean),
\ref{tab:DiffRec-amazon-book_clean_original-result-appendix} for Amazon-Books (clean),
\ref{tab:DiffRec-ml-1m_noisy_original-result-appendix} for MovieLens-1M (natural noise),
\ref{tab:DiffRec-yelp_noisy_original-result-appendix} for Yelp (natural noise), and
\ref{tab:DiffRec-amazon-book_noisy_original-result-appendix} for Amazon-Books (natural noise).

\hyperparamsArticleTable{DiffRec}{DiffRec}{0}{labeled ``(original hparams).''}
\hyperparamsArticleTable{L-DiffRec}{L-DiffRec}{0}{labeled ``(original hparams).''}
\hyperparamsArticleTable{T-DiffRec}{T-DiffRec}{1}{labeled ``(original hparams).''}
\hyperparamsArticleTable{LT-DiffRec}{LT-DiffRec}{1}{labeled ``(original hparams).''}

\begin{table}[th!]
    \caption{DiffRec (and variants) hyperparameter values modified in the experiments labeled ``(new hparams)'' with respect to the experiments labeled ``(original hparams).''}
    \label{tab:new-hparams-diffrec}
    \footnotesize
    \centering
    \begin{tabular}{lllc}
    \toprule
    Algorithm   & Dataset                       & Hyperparameter            & Value         \\    
    \midrule
    DiffRec     & MovieLens-1M (natural noise)  & Epochs                    & 300           \\
    \cmidrule(lr){2-4}
               & Yelp (natural noise)          & Epochs                    & 300           \\
               &                               & Reweight                  & True          \\
    \midrule
    L-DiffRec & MovieLens-1M (clean)          & Learning rate             & 0.001         \\
               &                               & Learning rate autoencoder          & 0.0005        \\
               &                               & VAE anneal cap            & 0             \\
    \midrule
    LT-DiffRec & Yelp (clean)                   & Epochs                    & 300           \\
               &                                & Batch size                & 400           \\
               &                                & VAE anneal steps          & 500           \\    
\bottomrule
\end{tabular}
\end{table}

\tableArticleResult{DiffRec}{DiffRec}{ml-1m_clean_original}{MovieLens-1M (clean)}{1}{}{1}
\tableArticleResult{DiffRec}{DiffRec}{yelp_clean_original}{Yelp (clean)}{1}{}{1}
\tableArticleResult{DiffRec}{DiffRec}{amazon-book_clean_original}{Amazon-Books (clean)}{1}{}{1}

\tableArticleResult{DiffRec}{DiffRec}{ml-1m_noisy_original}{MovieLens-1M (natural noise)}{1}{}{1}
\tableArticleResult{DiffRec}{DiffRec}{yelp_noisy_original}{Yelp (natural noise)}{1}{}{1}
\tableArticleResult{DiffRec}{DiffRec}{amazon-book_noisy_original}{Amazon-Books (natural noise)}{1}{
}{1}

\clearpage

\subsection{CF-Diff}
\label{app:cfdiff}

The hyperparameter values used in our experiments are reported in Table \ref{tab:CF-Diff-hyperparams} and the results for all the datasets and baseline algorithms are reported in Table
\ref{tab:CF-Diff-ML-1M_original-result-appendix} for MovieLens-1M (clean),
\ref{tab:CF-Diff-yelp2018_original-result-appendix} for Yelp (clean), and
\ref{tab:CF-Diff-anime_original-result-appendix} for Anime.

\hyperparamsArticleTable{CFDiff}{CF-Diff}{1}{}

\tableArticleResult{CFDiff}{CF-Diff}{ML-1M_original}{MovieLens-1M (clean)}{1}{
}{1}
\tableArticleResult{CFDiff}{CF-Diff}{yelp2018_original}{Yelp (clean)}{1}{}{1}
\tableArticleResult{CFDiff}{CF-Diff}{anime_original}{Anime}{1}{}{1}

\clearpage

\subsection{GiffCF}
\label{app:giffcf}
The hyperparameter values used in our experiments are reported in Table \ref{tab:GiffCF-hyperparams} and the results for all the datasets and baseline algorithms are reported in Table
\ref{tab:GiffCF-ml-1m_clean_original-result-appendix} for MovieLens-1M (clean),
\ref{tab:GiffCF-yelp_clean_original-result-appendix} for Yelp (clean), and
\ref{tab:GiffCF-amazon-book_clean_original-result-appendix} for Amazon-Books (clean).

\hyperparamsArticleTable{GiffCF}{GiffCF}{1}{}

\tableArticleResult{GiffCF}{GiffCF}{ml-1m_clean_original}{MovieLens-1M (clean)}{0}{}{1}
\tableArticleResult{GiffCF}{GiffCF}{yelp_clean_original}{Yelp (clean)}{0}{}{1}
\tableArticleResult{GiffCF}{GiffCF}{amazon-book_clean_original}{Amazon-Books (clean)}{0}{}{1}

\clearpage

\subsection{DDRM}
\label{app:ddrm}

The hyperparameters for experiments labeled “(original hparams)” are reported in Table \ref{tab:DDRM-hyperparams-common} (common fixed values) and Table \ref{tab:DDRM-hyperparams} (dataset- or backbone-specific values), recovered from the available artifacts. The hyperparameters for experiments labeled “(new hparams)” are reported in Table \ref{tab:new-hparams-ddrm}, as provided by the authors. Results for all datasets and baseline algorithms are reported in Table
\ref{tab:DDRM-ml-1m_noisy_original-result-appendix} for MovieLens-1M (natural noise),
\ref{tab:DDRM-yelp_noisy_original-result-appendix} for Yelp (natural noise), and
\ref{tab:DDRM-amazon-book_noisy_original-result-appendix} for Amazon-Books (natural noise).

\begin{table}[th!]
    \caption{DDRM common hyperparameter values used in all reproducibility experiments.}
    \label{tab:DDRM-hyperparams-common}
    \footnotesize
    \centering
    \begin{tabular}{ll|c}
\toprule
    Hyperparameter          & Described in          & Value         \\
\midrule
    Weight decay            & Source Code (default) & 0.0001        \\
    Batch size              & Source Code (default) & 2048          \\
    Mean type               & Source Code (default) & ``x0''          \\
    Noise schedule          & Source Code (default) & ``linear-var''  \\
    Sampling noise          & Source Code (default) & False         \\
    Time  embedding size    & Source Code (default) & 10            \\
    History num. per term   & Source Code (default) & 10            \\
    Beta fixed              & Source Code (default) & True          \\
    Normalize               & Source Code (default) & False         \\
    Dropout                 & Source Code (default) & 0.5           \\
    Latent dimension backbone & Paper (fixed)         & 64            \\
    If using dropout        & Source Code (default) & False         \\
    Keep prob               & Source Code (default) & 0.6           \\
    Dropout MLP             & Source Code (default) & 0.5           \\
    A n fold                & Source Code (default) & 100           \\
    A split                 & Source Code (default) & False         \\
    Dropout rate            & Source Code (default) & 0.2           \\
    Exponent                & Source Code (default) & 1             \\
    Num gradual             & Source Code (default) & 30,000        \\
\bottomrule
\end{tabular}
\end{table}

\begin{table}[th!]
    \caption{DDRM hyperparameter values used in the reproducibility experiments labeled ``(original hparams).''}
    \label{tab:DDRM-hyperparams}
    \footnotesize
    \centering
    \begin{tabular}{lll|ccc}
    \toprule
    Backbone & Hyperparameter          & Described in          & \begin{tabular}{c}MovieLens-1M\\(natural noise)\end{tabular}     & \begin{tabular}{c}Yelp\\(natural noise)\end{tabular}              & \begin{tabular}{c}Amazon-Books\\(natural noise)\end{tabular}           \\    
    \midrule
    LightGCN & Epochs         & Source Code & 1000          & 1000              & 1000      \\
    & MLP learning rate                  & Source Code & 0.00001       & 0.00001           & 0.00001   \\
    & Diffusion learning rate            & Source Code & 0.001         & 0.001             & 0.001     \\
    & Test batch size         & -  & 1000          & 1000                 & 500       \\
    & Noise Scale             & Source Code & 0.01          & 0.0001            & 0.01  \\
    & Noise max               & Source Code & 0.001         & 0.01              & 0.001  \\
    & Noise min               & Source Code & 0.0001        & 0.0001            & 0.0001  \\
    & Diffusion steps         & Source Code & 50            & 30                & 50  \\
    & Sampling steps          & Source Code & 62            & 37                & 62  \\
    & MLP dims                & Source Code & ``[200,1100]''  & ``[200,1000]''      & ``[200,1000]''  \\
    & Num layers              & Source Code & 2             & 3                 & 2  \\
    & Activation              & Source Code & ``relu''        & ``relu''            & ``relu''  \\
    & Alpha                & Source Code & 0.1           & 0.3               & 0.5  \\
    & Beta                 & Source Code & 0             & 0.8               & 0.0  \\
\midrule
    MF & Epochs         & Source Code & 1000            & 1000              & 1000  \\
    & MLP learning rate                  & Source Code & 0.00001   & 0.00001           & 0.00001  \\
    & Diffusion learning rate            & Source Code & 0.001     & 0.001             & 0.001 \\
    & Test batch size         & -  & 1000          & 1000                 & 500       \\
    & Noise Scale             & Source Code & 0.0001    & 0.0001            & 0.001  \\
    & Noise max               & Source Code & 0.001     & 0.1               & 0.001  \\
    & Noise min               & Source Code & 0.0001    & 0.001             & 0.0001  \\
    & Diffusion steps         & Source Code & 20        & 50                & 3  \\
    & Sampling steps          & Source Code & 20        & 50                & 3  \\
    & MLP dims                & Source Code & ``[100,1000]'' & ``[200,1000]''   & ``[200,600]''  \\
    & Num layers              & Source Code & 4         & 2                 & 0  \\
    & Activation              & Source Code & ``sigmoid''  & ``relu''           & ``sigmoid''  \\
    & Alpha                & Source Code & 0.4       & 0.2               & 0.2  \\
    & Beta                 & Source Code & 1.0       & 0.9               & 0.9  \\
\midrule
    SGL & Epochs            & Source Code & 1000        & 1000              & 1000         \\
    & MLP learning rate                  & Source Code & 0.001     & 0.00001           & 0.00001  \\
    & Diffusion learning rate            & Source Code & 0.00001   & 0.001             & 0.001  \\
    & Test batch size         & -  & 1000          & 1000                 & 500       \\
    & Noise scale             & Source Code & 0.001     & 0.0001            & 0.0001  \\
    & Noise max               & Source Code & 0.001     & 0.01              & 0.001  \\
    & Noise min               & Source Code & 0.0001    & 0.001             & 0.0001  \\
    & Diffusion steps         & Source Code & 50        & 50                & 3  \\
    & Sampling steps          & Source Code & 50        & 62                & 3  \\
    & MLP dims                & Source Code & ``[300,1000]'' & ``[200,1000]''   & ``[300,1000]''  \\
    & Num layers              & Source Code & 5         & 2                 & 2  \\
    & Activation              & Source Code & ``sigmoid'' & ``sigmoid''         & ``sigmoid''  \\
    & Alpha                & Source Code & 0.2       & 0.3               & 0.6  \\
    & Beta                 & Source Code & 0.3       & 0.05              & 0.5  \\
\bottomrule
\end{tabular}
\end{table}

\begin{table}[th!]
    \caption{DDRM hyperparameter values modified in the experiments labeled ``(new hparams)'' with respect to the experiments labeled ``(original hparams).''}
    \label{tab:new-hparams-ddrm}
    \footnotesize
    \centering
    \begin{tabular}{llcc}
    \toprule
    Algorithm       & Dataset                       & Hyperparameter            & Value         \\
    \midrule
    DDRM-LightGCN   & MovieLens-1M (natural noise)  & Test batch size           & 512           \\
                    & Yelp (natural noise)          & Test batch size           & 512           \\
                    & Amazon-Books (natural noise)  & Test batch size           & 128           \\
    \midrule
    DDRM-MF         & MovieLens-1M (natural noise)  & Test batch size           & 512           \\
                    &                               & Sampling steps            & 25            \\
                    &                               & MLP dims                  & [200,1200]    \\
                    &                               & Num layers                & 3             \\
                    \cmidrule(lr){2-4}
                    & Yelp (natural noise)          & Test batch size           & 512           \\
                    &                               & Noise scale               & 0.001         \\
                    &                               & Noise max                 & 0.01          \\
                    &                               & Noise min                 & 0.0001        \\
                    &                               & Num layers                & 3             \\
                    &                               & Activation                       & sigmoid       \\
                    &                               & Beta                      & 0.95          \\
                    \cmidrule(lr){2-4}
                    & Amazon-Books (natural noise)  & Test batch size           & 128           \\
    \midrule
    DDRM-SGL        & MovieLens-1M (natural noise)  & Test batch size           & 256           \\
                    &                               & Noise scale               & 0.0001        \\
                    &                               & Diffusion steps           & 30            \\
                    &                               & Sampling steps            & 37            \\
                    \cmidrule(lr){2-4}
                    & Yelp (natural noise)          & Test batch size           & 256           \\
                    &                               & Sampling steps            & 50            \\
                    \cmidrule(lr){2-4}
                    & Amazon-Books (natural noise)  & Test batch size           & 128           \\
\bottomrule
\end{tabular}
\end{table}

\tableArticleResult{DDRM}{DDRM}{ml-1m_noisy_original}{MovieLens-1M (natural noise)}{0}{}{1}
\tableArticleResult{DDRM}{DDRM}{yelp_noisy_original}{Yelp (natural noise)}{0}{}{1}
\tableArticleResult{DDRM}{DDRM}{amazon-book_noisy_original}{Amazon-Books (natural noise)}{0}{}{1}

\clearpage

\section{Baseline Hyperparameter Ranges}
\label{app:baseline-hparams}
In this section we report the hyperparameter ranges and distribution for all the baselines in our experiments, see Table \ref{tab:hyperparameters_our_baselines-KNN} (Nearest-Neighbor),
\ref{tab:hyperparameters_our_baselines-graph_based} (Graph-based),
\ref{tab:hyperparameters_our_baselines-IB_ML} (Item-based Machine Learning),
\ref{tab:hyperparameters_our_baselines-MF} (Matrix Factorization), and
\ref{tab:hyperparameters_our_baselines-AE} (Autoencoder).

\makeatletter
\newcommand\footnoteref[1]{\protected@xdef\@thefnmark{\ref{#1}}\@footnotemark}
\makeatother

\begin{table}[h]
    \begin{minipage}{\textwidth}
    \footnotesize
    \centering
    \begin{tabular}{cl|cccc}
    \toprule
    Algorithm	& Hyperparameter	&  Range	 & Type     & Distribution	\\
    \midrule
    \multirow{5}{*}{\begin{tabular}{c}UserKNN, ItemKNN \end{tabular}}  	
    				&topK	        & 5 - 1000 	& Integer   & uniform 	\\
    				&shrink	        & 0 - 1000 	& Integer   & uniform 	\\
    				&similarity	    & cosine 	& Categorical 	& 	\\
    				&normalize\footnote{\label{foot:knn_normalize}The \emph{normalize} hyperparameter in KNNs refers to the use of the denominator when computing the similarity.} 	    & True, False 	& Categorical 	& 	\\
    				&feature weighting	& none, TF-IDF, BM25 	& Categorical 	& 	\\
	\bottomrule
   	\end{tabular}
   	\end{minipage}
    \caption{Hyperparameter ranges and distributions for our Nearest-Neighbor baselines.}
    \label{tab:hyperparameters_our_baselines-KNN}
\end{table}

\begin{table}[h]
    \begin{minipage}{\textwidth}
    \footnotesize
    \centering
    \begin{tabular}{cl|cccc}
    \toprule
    Algorithm	& Hyperparameter	&  Range	 & Type     & Distribution	\\
    \midrule
    \multirow{3}{*}{\palpha}  	
     				&topK	        & 5 - 1000 	& Integer   & uniform 	\\
     				&alpha	        & 0 - 2	& Real   & uniform 	\\
     				&normalize similarity\footnote{\label{foot:normalize_similarity}The \emph{normalize similarity} hyperparameter refers to applying L1 regularization on the rows of the similarity matrix.}	& True, False 	& Categorical 	& 	\\
    \midrule
    \multirow{4}{*}{\pbeta}  	
    				&topK	        & 5 - 1000 	& Integer   & uniform 	\\
    				&alpha	        & 0 - 2	& Real   & uniform 	\\
    				&beta	        & 0 - 2	& Real   & uniform 	\\
    				&normalize similarity\footnoteref{foot:normalize_similarity}	& True, False 	& Categorical 	& 	\\
    \midrule
    \multirow{3}{*}{GF-CF}  	
    				&alpha	        &$10^{-3}$ - $10^{+3}$     & Real   & log-uniform 	\\
    				&num factors	& 1 - 350  	 & Integer   & uniform 	\\
    \midrule
    \multirow{3}{*}{LightGCN}  	
    				&epochs	        & 1 - 1000 	 & Integer   & early-stopping 	\\
                    &GNN layers K	& 1 - 6 	 & Integer   & uniform 	\\
                    &embedding size & 2 - 350 	 & Integer   & uniform 	\\
    				&batch size	    & 256, 512, 1024, 2048, 4096     & Categorical   &  	\\
    				&learning rate  & $10^{-6}$ - $10^{-1}$     & Real   & log-uniform 	\\
                    &l2 reg         & $0.1$ - $0.9$     & Real   & uniform 	\\
                    &dropout rate   & $10^{-6}$ - $10^{-1}$     & Real   & log-uniform 	\\
                    &sgd mode	    & sgd, adam, adagrad, rmsprop    & Categorical 	& 	\\
	\bottomrule
   	\end{tabular}
   	\end{minipage}
    \caption{Hyperparameter ranges and distributions for our Graph-based baselines.}
    \label{tab:hyperparameters_our_baselines-graph_based}
\end{table}

\begin{table}[h]
    \begin{minipage}{\textwidth}
    \footnotesize
    \centering
    \begin{tabular}{ll|ccl}
    \toprule
    Algorithm	& Hyperparameter	&  Range	 & Type     & Distribution	\\
    \midrule
    \multirow{1}{*}{\EASER}  	
    				&l2 norm	& $10^{0}$ - $10^{+7}$     & Real   & log-uniform 	\\
                    & normalize matrix & True, False      & Categorical   & \\
                    & topK      & None                      & Categorical & \\
    \midrule
    \multirow{3}{*}{SLIM}  	
    				&topK	        & 5 - 1000 	& Integer   & uniform 	\\
    				&l1 ratio	    & $10^{-5}$ - $10^{0}$     & Real   & log-uniform 	\\
    				&alpha	        & $10^{-3}$ - $10^{0}$     & Real   & uniform 	\\
    \midrule
    \multirow{7}{*}{SLIM-BPR }  	
     				&topK	        & 5 - 1000 	& Integer   & uniform 	\\
     				&epochs	        & 1 - 1500 	& Integer 	& early-stopping 	\\
     				&symmetric	    & True, False 	& Categorical 	& 	\\
     				&sgd mode	    & sgd, adam, adagrad    & Categorical 	& 	\\
     				&lambda i   	& $10^{-5}$ - $10^{-2}$     & Real   & log-uniform 	\\
     				&lambda j   	& $10^{-5}$ - $10^{-2}$     & Real   & log-uniform 	\\
     				&learning rate 	& $10^{-4}$ - $10^{-1}$     & Real   & log-uniform 	\\
	\bottomrule
   	\end{tabular}
   	\end{minipage}
    \caption{Hyperparameter ranges and distributions for our Item-based Machine Learning baselines.}
    \label{tab:hyperparameters_our_baselines-IB_ML}
\end{table}

\begin{table}[h]
    \begin{minipage}{\textwidth}
    \footnotesize
    \centering
    \begin{tabular}{ll|ccl}
    \toprule
    Algorithm	& Hyperparameter	&  Range	 & Type     & Distribution	\\
    \midrule
    \multirow{10}{*}{MultVAE}  	
                &epochs	        & 1 - 500\footnote{\label{foot:epochs_lower_slow}The number of epochs is lower due to the algorithm being slower, but converging in a lower number of epochs.} 	    & Integer 	& early-stopping 	\\
                &learning rate  & $10^{-6}$ - $10^{-2}$     & Real   & log-uniform 	\\
                &l2 reg         & $10^{-6}$ - $10^{-2}$     & Real   & log-uniform 	\\
                &dropout        & $0$ - $0.8$     & Real   & uniform 	\\
                &annealing steps  & $10^{5}$ - $6\cdot 10^{5}$     & Integer   & uniform \\
                &anneal cap     & $0$ - $0.6$     & Real   & uniform 	\\
                &batch size	    & 128, 256, 512, 1024                & Categorical 	& 	\\
                &encoding size  & 1 - 512     & Integer   & uniform 	\\
                &layer size multiplier\footnote{This hyperparameter is used to generate the decoder architecture. Starting from the encoding size the size of the next hidden layer is computed as the product of the previous one and the layer multiplier. The process terminates when either the desired number of hidden layers is reached or any further hidden layer added would exceed the size of the input data.}  & 2 - 10     & Integer   & uniform 	\\
                &max n hidden layers    & 2 - 4     & Integer   & uniform 	\\
	\bottomrule
   	\end{tabular}
   	\end{minipage}
    \caption{Hyperparameter ranges and distributions for our Autoencoder baselines.}
    \label{tab:hyperparameters_our_baselines-AE}
\end{table}

\begin{table}[h]
    \begin{minipage}{\textwidth}
    \footnotesize
    \centering
    \begin{tabular}{ll|ccl}
    \toprule
    Algorithm	& Hyperparameter	&  Range	 & Type     & Distribution	\\
    \midrule
    \multirow{7}{*}{MF-BPR}  	
    				&num factors	& 1 - 200\footnoteref{foot:num_factor_lower_slow}  	& Integer   & uniform 	\\
    				&epochs	        & 1 - 1500 	& Integer 	& early-stopping 	\\
    				&sgd mode	    & sgd, adam, adagrad    & Categorical 	& 	\\
    				&batch size	    & $1, 2, 4, 8, 16, 32, 64, 128, 256, 512, 1024$    & Categorical 	& 	\\
    				&positive reg  	& $10^{-5}$ - $10^{-2}$     & Real   & log-uniform 	\\
    				&negative reg  	& $10^{-5}$ - $10^{-2}$     & Real   & log-uniform 	\\
    				&learning rate 	& $10^{-4}$ - $10^{-1}$     & Real   & log-uniform 	\\
    \midrule
    \multirow{7}{*}{MF-WARP}  	
    				&num factors	& 1 - 200\footnoteref{foot:num_factor_lower_slow}  	& Integer   & uniform 	\\
    				&epochs	        & 1 - 1500 	& Integer 	& early-stopping 	\\
    				&sgd mode	    & sgd, adam, adagrad    & Categorical 	& 	\\
    				&batch size	    & $1, 2, 4, 8, 16, 32, 64, 128, 256, 512, 1024$    & Categorical 	& 	\\
    				&positive reg  	& $10^{-5}$ - $10^{-2}$     & Real   & log-uniform 	\\
    				&negative reg  	& $10^{-5}$ - $10^{-2}$     & Real   & log-uniform 	\\
    				&learning rate 	& $10^{-4}$ - $10^{-1}$     & Real   & log-uniform 	\\
                    &neg item attempts 	& 5, 10, 15, 20     & Categorical   & \\
     \midrule
     \multirow{9}{*}{SVDpp}  	
                    &num factors	& 1 - 200\footnote{\label{foot:num_factor_lower_slow}The number of factors is lower than PureSVD due to the algorithm being slower.} 	& Integer   & uniform 	\\
                    &epochs	        & 1 - 500\footnote{\label{foot:epochs_lower_slow}The number of epochs is lower than SLIM-BPR or MF-BPR due to the algorithm being slower.} 	& Integer 	& early-stopping 	\\
                    &use bias	    & True, False 	& Categorical 	& 	\\
    				&sgd mode	    & sgd, adam, adagrad    & Categorical 	& 	\\
    				&batch size	    & $1, 2, 4, 8, 16, 32, 64, 128, 256, 512, 1024$    & Categorical 	& 	\\
    				&item reg   	& $10^{-5}$ - $10^{-2}$     & Real   & log-uniform 	\\
    				&user reg   	& $10^{-5}$ - $10^{-2}$     & Real   & log-uniform 	\\
    				&learning rate  & $10^{-4}$ - $10^{-1}$     & Real   & log-uniform 	\\
    				&negative quota\footnote{The \emph{negative quota} is the percentage of samples chosen among items unobserved by the user, having a target rating of 0.}	& 0.00 - 0.50 	& Real   & uniform 	\\
     \midrule
     \multirow{1}{*}{PureSVD}  	
     				&num factors	& 1 - 350 	& Integer   & uniform 	\\
    \midrule
    \multirow{6}{*}{\iALS}  	
    				&num factors    & 1 - 200\footnote{\label{foot:num_factor_lower_slow}The number of factors is lower due to the algorithm being slower.} 	    & Integer       & uniform 	\\
    				&epochs	        & 1 - 300\footnote{\label{foot:epochs_lower_slow}The number of epochs is lower due to the algorithm being slower, but converging in a lower number of epochs.} 	& Integer 	& early-stopping 	\\
    				&confidence scaling	&linear, log    & Categorical 	& 	\\
    				&alpha	    & $10^{-3}$ - $50$ \footnote{\label{foot:hyperparameter_value_original_article}The maximum value of this hyperparameter had been suggested in the article proposing the algorithm.}      & Real   & log-uniform 	\\
    				&epsilon	& $10^{-3}$ - $10^{+1}$ \footnoteref{foot:hyperparameter_value_original_article}     & Real   & log-uniform 	\\
    				&reg	    & $10^{-5}$ - $10^{-2}$     & Real   & log-uniform 	\\
	\bottomrule
   	\end{tabular}
   	\end{minipage}
    \caption{Hyperparameter ranges and distributions for our Matrix Factorization baselines.}
    \label{tab:hyperparameters_our_baselines-MF}
\end{table}

\end{document}